\documentclass[a4paper,fleqn]{cas-dc}

\usepackage[numbers]{natbib}
\usepackage{threeparttablex, booktabs, url, amsmath, xr, flushend} %balance
\usepackage{orcidlink}
\usepackage{subfiles}

\def\tsc#1{\csdef{#1}{\textsc{\lowercase{#1}}\xspace}}
\tsc{WGM}
\tsc{QE}

\newcommand{\petar}{\texttt{PETAR}}
\begin{document}
% Short title
\shorttitle{Escaping Pulsar from an Ancient Open Cluster}
% Short author
\shortauthors{Zhang et al.}

% Main title of the paper
\title [mode = title]{A pulsar escaping an ancient open cluster via tidal stripping}

\author[1,2,3]{Lei Zhang\orcidlink{0000-0001-8539-4237}} 
\author[4]{Xiaoting Fu\orcidlink{0000-0002-6506-1985}}\cormark[1]\ead{xiaoting.fu@pmo.ac.cn}
\author[3,1]{Di Li\orcidlink{0000-0003-3010-7661}}\cormark[1]\ead{dili@tsinghua.edu.cn}
\author[2,5]{Emma Carli\orcidlink{0000-0003-3265-2866}}
\author[6,7]{Long Wang\orcidlink{0000-0001-8713-0366}}
\author[8,3]{Jiaqi Zhao\orcidlink{0000-0002-7716-1166}}
\author[8]{Craig O. Heinke\orcidlink{0000-0003-3944-6109}}
\author[9]{Emanuele Dalessandro\orcidlink{0000-0003-4237-4601}}
\author[10]{Yang Chen\orcidlink{0000-0002-3759-1487}}
\author[9]{Angela Bragaglia\orcidlink{0000-0002-0338-7883}}
\author[11,12]{Duncan R. Lorimer\orcidlink{0000-0003-1301-966X}}
\author[13]{Ewan D. Barr\orcidlink{0000-0001-8715-9628}}
\author[14]{Sarah Buchner}
\author[15]{Shi Dai\orcidlink{0000-0003-3294-3081}}
\author[16,17]{Zhiyu Zhang\orcidlink{0000-0003-3294-3081}}
\author[18]{Erbil G\"{u}gercino\u{g}lu}
\author[19]{Alessandro Ridolfi\orcidlink{0000-0001-6762-2638}}
\author[20]{Jie Zhang}
\author[21,22]{Meng Guo}
\author[22]{Mengmeng Ni}
\author[22]{Jiale Hu}
\author[23,24]{Yi Feng\orcidlink{0000-0001-6762-2638}}
\author[1]{Pei Wang\orcidlink{0000-0002-3386-7159}}
\author[25]{Qijun Zhi\orcidlink{0000-0001-9389-5197}}

% Address/affiliation
\affiliation[1]{organization={State Key Laboratory of Radio Astronomy and Technology, National Astronomical Observatories, Chinese Academy of Sciences},
            city={Beijing},
            citysep={}, % Uncomment if no comma needed between city and postcode
            postcode={100101}, 
            country={China}}
            
\affiliation[2]{organization={Centre for Astrophysics and Supercomputing, Swinburne University of Technology},
            city={Melbourne},
            citysep={}, % Uncomment if no comma needed between city and postcode
            postcode={3122},
            country={Australia}}
            
\affiliation[3]{organization={New Cornerstone Laboratory, Department of Astronomy},
            addressline={Tsinghua University}, 
            city={Beijing},
            citysep={}, % Uncomment if no comma needed between city and postcode
            postcode={100084}, 
            country={China}}   
            
\affiliation[4]{organization={Purple Mountain Observatory},
            addressline={Chinese Academy of Sciences}, 
            city={Nanjing},
            citysep={}, % Uncomment if no comma needed between city and postcode
            postcode={210023}, 
            country={China}}

\affiliation[5]{organization={ARC Centre of Excellence for Gravitational Wave Discovery (OzGrav), Swinburne University of Technology},
            city={Melbourne},
            citysep={}, % Uncomment if no comma needed between city and postcode
            postcode={3122},
            country={Australia}}
            
\affiliation[6]{organization={School of Physics and Astronomy},
            addressline={Sun Yat-sen University}, 
            city={Zhuhai},
            citysep={}, % Uncomment if no comma needed between city and postcode
            postcode={519082}, 
            country={China}}             

\affiliation[7]{organization={CSST Science Center for the Guangdong-Hong Kong-Macau Greater Bay Area},
            city={Zhuhai},
            citysep={}, % Uncomment if no comma needed between city and postcode
            postcode={519082}, 
            country={China}}

\affiliation[8]{organization={Department of Physics, CCIS 4-183},
            addressline={University of Alberta}, 
            city={Edmonto},
            citysep={}, % Uncomment if no comma needed between city and postcode
            postcode={AB T6G 2E1}, 
            country={Canada}} 
            
\affiliation[9]{organization={INAF – Astrophysics and Space Science Observatory of Bologna},
            city={Bologna},
            citysep={ }, % Uncomment if no comma needed between city and postcode
            postcode={40129},
            country={Italy}}

\affiliation[10]{organization={School of Physics and Optoelectronic Engineering},
            addressline={Anhui University}, 
            city={Hefei},
            citysep={}, % Uncomment if no comma needed between city and postcode
            postcode={230601}, 
            country={China}} 

\affiliation[11]{organization={Center for Gravitational Waves and Cosmology, West Virginia University},
            city={Morgantown},
            citysep={}, % Uncomment if no comma needed between city and postcode
            postcode={26506},
            country={USA}}
 
\affiliation[12]{organization={Department of Physics and Astronomy, West Virginia University},
            city={Morgantown},
            citysep={}, % Uncomment if no comma needed between city and postcode
            postcode={26506},
            country={USA}}   

\affiliation[13]{organization={Max-Planck-Institut f\"{u}r Radioastronomie},
            city={Bonn},
            citysep={}, % Uncomment if no comma needed between city and postcode
            postcode={53121},
            country={Germany}}
            
\affiliation[14]{organization={South African Radio Astronomy Observatory},
            city={Cape Town},
            citysep={}, % Uncomment if no comma needed between city and postcode
            postcode={7925},
            country={South Africa}}        
 
\affiliation[15]{organization={CSIRO Astronomy and Space Science},
            city={Epping},
            citysep={}, % Uncomment if no comma needed between city and postcode
            postcode={1710},
            country={Australia}}
                       
\affiliation[16]{organization={School of Astronomy and Space Science},
            addressline={Nanjing University}, 
            city={Nanjing},
            citysep={}, % Uncomment if no comma needed between city and postcode
            postcode={210023}, 
            country={China}} 
            
\affiliation[17]{organization={Key Laboratory of Modern Astronomy and Astrophysics},
            addressline={Nanjing University, Ministry of Education}, 
            city={Nanjing},
            citysep={}, % Uncomment if no comma needed between city and postcode
            postcode={210023}, 
            country={China}}             
            
\affiliation[18]{organization={School of Arts and Sciences, Qingdao Binhai University},
            city={Qingdao},
            citysep={}, % Uncomment if no comma needed between city and postcode
            postcode={266555},
            country={China}}
            
\affiliation[19]{organization={Fakult\"at f\"ur Physik, Universit\"at Bielefeld},
            city={Bielefeld},
            citysep={}, % Uncomment if no comma needed between city and postcode
            postcode={33501},
            country={Germany}}
            
\affiliation[20]{organization={School of Arts and Sciences},
            addressline={Shanghai Dianji University}, 
            city={Shanghai},
            citysep={}, % Uncomment if no comma needed between city and postcode
            postcode={200240}, 
            country={China}}             

\affiliation[21]{organization={National Supercomputing Center in Jinan},
            addressline={Qilu University of Technology}, 
            city={Jinan},
            citysep={}, % Uncomment if no comma needed between city and postcode
            postcode={250103}, 
            country={China}}                        

\affiliation[22]{organization={Jinan Institute of Supercomputing Technology},
            addressline={28666 East Jingshi Road}, 
            city={Jinan},
            citysep={}, % Uncomment if no comma needed between city and postcode
            postcode={250103}, 
            country={China}} 

\affiliation[23]{organization={Research Center for Intelligent Computing Platforms},
            addressline={Zhejiang Laboratory}, 
            city={Hangzhou},
            citysep={}, % Uncomment if no comma needed between city and postcode
            postcode={311100}, 
            country={China}}     

\affiliation[24]{organization={Institute for Astronomy, School of Physics},
            addressline={Zhejiang University}, 
            city={Hangzhou},
            citysep={}, % Uncomment if no comma needed between city and postcode
            postcode={310027}, 
            country={China}}                     
            
\affiliation[25]{organization={School of Physics},
            addressline={Guizhou University}, 
            city={Guiyang},
            citysep={}, % Uncomment if no comma needed between city and postcode
            postcode={550025}, 
            country={China}}

% Corresponding author text
\cortext[1]{Corresponding author}
% Footnote text
%\fntext[1]{}
%\fntext[fn1]{These authors contributed equally: XX, XX}

%\iffalse
\begin{abstract}
-
%Open clusters are the primary birthplaces of stars in the Milky Way disk, yet their neutron star progeny are rarely found within them, presumably due to supernova-induced kicks that eject them at birth. Here we report the arcsec-level localization of the pulsar PSR~J1921$+$3745 to the tidal tail of NGC 6791, one of the oldest and most massive open clusters. Our N-body simulation shows that more than 95\% of neutron stars formed in such clusters have been ejected.  This pulsar’s location in the tidal tail indicates it was retained for billions of years before being stripped by Galactic tides. This long-term retention requires low natal kicks, consistent with formation via electron-capture supernova. Our findings capture a rare snapshot of a neutron star transitioning into the Galactic field, identifying tidal stripping of ancient clusters as a verified source of the Galactic neutron star population.
\end{abstract}
%\fi
% For a title note without a number/mark
%\keywordtitle { } { Keywords }
%\nonumnote{}
%SB article: Normally including a 250-word abstract, 4–6 keywords, up to 6 figures or tables and 60 references.
%SB Short Communications 3 pages, 1-2 displayed items , <15 references
%\begin{keywords}
%Radio pulsars \sep Neutron stars \sep Open clusters
%\end{keywords}

\maketitle

%%%############### Introduction ############### 
Most stars are initially formed within star clusters.% \cite{Lada2003}.
These stellar systems contribute to the Galactic field population through the escape of individual stars and the gradual dissolution of the clusters over cosmic time. In the Milky Way, the majority of stars in the Galactic disk are thought to have originated in Open Clusters (OCs), whereas halo stars are understood to originate mostly from accreted dwarf galaxies and disrupted globular clusters \cite{Horta2021}. 

Neutron stars (NSs), being the remnants of massive stars, naturally inherit this cluster origin. However, their subsequent retention within or escape from their parent clusters provides a crucial link between small-scale supernova physics and large-scale Galactic tidal interactions.
The removal of  NSs from OCs is primarily driven by two pathways: tidal interactions with the Galactic gravitational potential and kicks imparted by supernova explosions during their formation. Tidal interactions gradually strip NSs (together with other stars) from clusters as they orbit around the Galaxy, whereas natal kicks from supernovae events can eject NSs effectively \cite{Lai2001}. 

Different formation channels impart distinct natal kick velocities. Core-collapse supernovae (CCSNe) can generate kicks often exceeding a few hundred km s$^{-1}$ \cite{Hobbs2005, Verbunt2017}, typically much larger than the escape velocities of star clusters ($\sim$1--90 km s$^{-1}$)\footnote{\url{https://people.smp.uq.edu.au/HolgerBaumgardt/globular/parameter.html}}. In contrast, NSs formed through low-kick channels, such as electron-capture supernovae (ECSNe), either via single star or binary channels \cite{Wang2026}, are expected to receive typically several km s$^{-1}$ \cite{Gessner2018} kicks, allowing them to remain bound and gradually escape through tidal interactions. Tidal tails of OCs, shaped by long-term evaporation and Galactic tides, are therefore expected to trace the leakage of such low-kick NSs into the Galactic field. However, despite this expectation, no neutron star has been robustly identified in the tidal tail of any open cluster to date.

NGC~6791 is the oldest ($\sim$8\,Gyr)~\cite{Brogaard2012} and one of the most massive OCs known in the Milky Way \cite{Hunt2024}. Given its unusually high mass for an OC and advanced age, NGC~6791 is expected to have produced a substantial population of NSs and undergone significant dynamical evolution. Its pronounced tidal distortion \cite{Dalessandro2015}, therefore making NGC~6791 a prime target for searching for cluster-born NSs that either remain in the bound cluster or are escaping along its tidal tails.

%%%############### Observations ###############
%\section{Observations and Localization}\label{sec:data}
Using the Five-hundred-meter Aperture Spherical radio Telescope (FAST~\cite{Nan2011,Li2018}) 19-beam receiver at L-band (1050–1450\,MHz), we conducted a targeted deep pulsar-search observation of NGC 6791, resulting in the discovery of PSR~J1921$+$3745 (reported previously as PSR J1922+37 in \cite{LiuXJ2025}), an isolated pulsar with a spin period of 1.9\,seconds and a dispersion measure (DM) of 86\,pc cm$^{-3}$. 
To precisely localize the pulsar, we performed a follow-up observation with the MeerKAT radio telescope, using the FBFUSE and APSUSE backends at UHF-band (544–1088 MHz). With a two-hour integration, we obtained a multi-beam detection of the pulsar, with signal to noise ratio S/N  $\sim$20. 

\begin{figure*}
    \centering
    \includegraphics[width=0.9\linewidth]{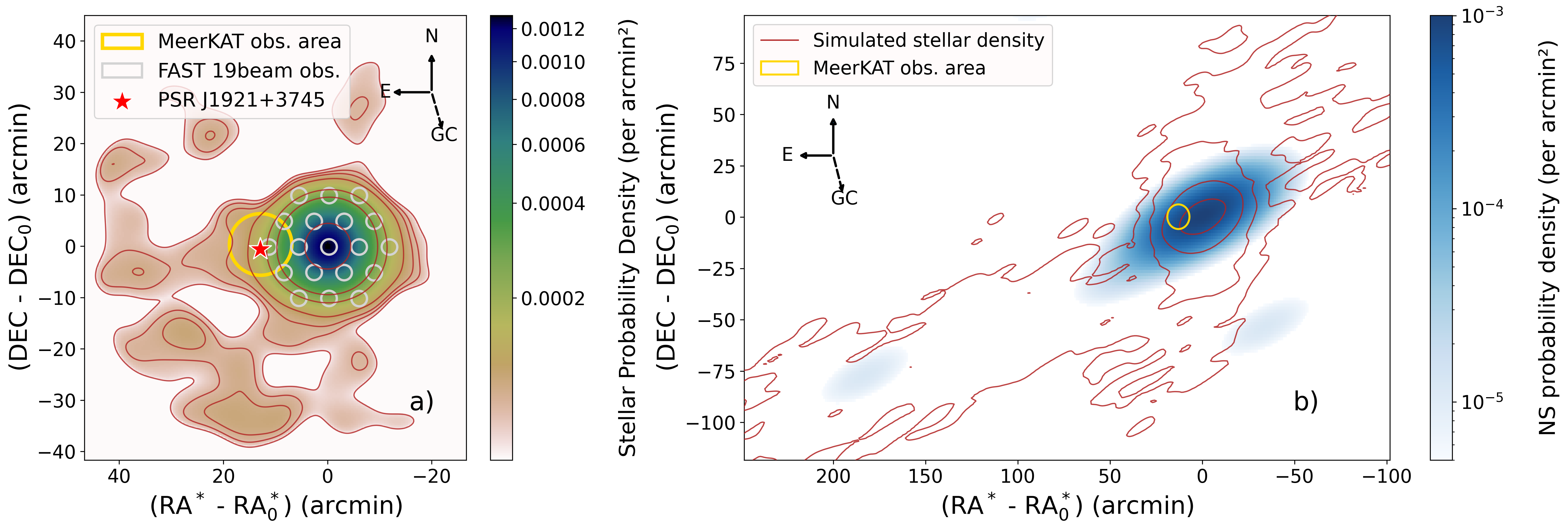}
    \caption{{\bf Stellar surface density map of NGC 6791 members from observation (panel a on the left) and simulation (panel b on the right).} The red star in panel a marks the position of PSR~J1921$+$3745. Grey and yellow circles in panel a delineate the FAST and MeerKAT search areas, respectively. 
    In panel b, red contours show the simulated stellar surface density at an age of 8 Gyr, while the color map shows the normalized probability density of simulated NSs.
    The projected coordinates are relative to the cluster center. Directions to the north, east, and the Galactic center are illustrated with arrows.}
\label{fig:radec_2fig}
\end{figure*}

The MeerKAT observations refine the pulsar’s position to (RA, Dec) = ($19^{\mathrm{h}}\,21^{\mathrm{m}}\,59^{\mathrm{s}}.70$, $+37^{\circ}\,45'\,50''.3$; J2000), with a 2-$\sigma$ uncertainty of ($0.084^{\mathrm{s}}, 3.3''$; Supplementary material
Section A.2). This position lies $\sim1.7$ arcmin, more than half the FAST beam width, from the center of our discovery beam and differs by $\sim0.75$ arcmin from the best-fit location reported by \cite{LiuXJ2025}, who independently identified the same pulsar. The MeerKAT position is also in good agreement with the subsequently reported timing position of \citet{LiuXJ2026}. 
Based on the deep photometry of NGC 6791 obtained with the MegaCAM on the Canada-France-Hawaii Telescope (CFHT) \cite{Dalessandro2015}, wherein main-sequence stars down to g$\sim$22.5 mag and out to $\sim$3000 arcsec from the cluster center are carefully selected, our analysis shows that PSR~J1921$+$3745 is located within a prominent tidal tail on the east side of the cluster (Fig. \ref{fig:radec_2fig}a). 

In addition to the stable integrated profile, sporadic single pulses are detected with peak intensities reaching 1–2$~\times~10^{3}$ times the average emission, indicating intermittent magnetospheric variability. Such brightness enhancements are comparable to those observed in giant pulses from energetic pulsars (see Supplementary material Section G). This behavior reflects magnetospheric variability but does not uniquely indicate a young evolutionary state.

%%%############### Cluster association ###############
%\section{Cluster association}\label{sec:association}
The source shows no persistent radio, X-ray, or optical counterpart, consistent with an isolated, non-accreting NS. 
The association of PSR~J1921$+$3745 with NGC~6791 is supported by three independent lines of evidence, including its distance consistency, low probability of chance alignment, and the absence of signatures of a recently formed NS. 
The dispersion-measure (DM) inferred distance, $d_{\rm DM} = 6.82 \pm 4.02$ kpc, is consistent within uncertainties with the cluster distance of $\sim$4 kpc derived from Gaia-based analyses and eclipsing binaries. 
The probability of a chance alignment with an unrelated field pulsar is low. Adopting pulsar population models, we estimate a surface density of $\sim$0.0096 deg$^{-2}$, corresponding to a probability of only $\sim$0.6\% over the area encompassing the cluster and $\sim$0.03\% in the MeerKAT observation area.
A more conservative estimation of $2\pm1\%$ chance alignment in the cluster area is discussed in the Supplementary Section D, in any case remains low.
An alternative interpretation as a recently formed neutron star is disfavored by the absence of nearby star-forming regions or young stellar populations. 
We therefore conclude that PSR~J1921$+$3745 is most likely physically associated with NGC~6791, although the evolutionary pathway leading to its present-day radio-emitting state remains uncertain (see Supplementary material Section E for details).

%%%############### NS retention ###############
%\section{Survival and stripping of neutron stars}\label{sec:NSretention}
To investigate the origin of PSR~J1921$+$3745, we performed direct $N$-body simulations with \petar~\cite{petar} tailored to the properties of NGC~6791 \cite{Fu2022}. The simulations include NS formation through both CCSNe and ECSNe with distinct natal kick distributions. By the cluster age of 8 Gyr, the model produces 973 NSs: 877 from CCSNe and 96 from ECSNe. 

\begin{figure*}
    \centering
    \includegraphics[width=0.8\linewidth]{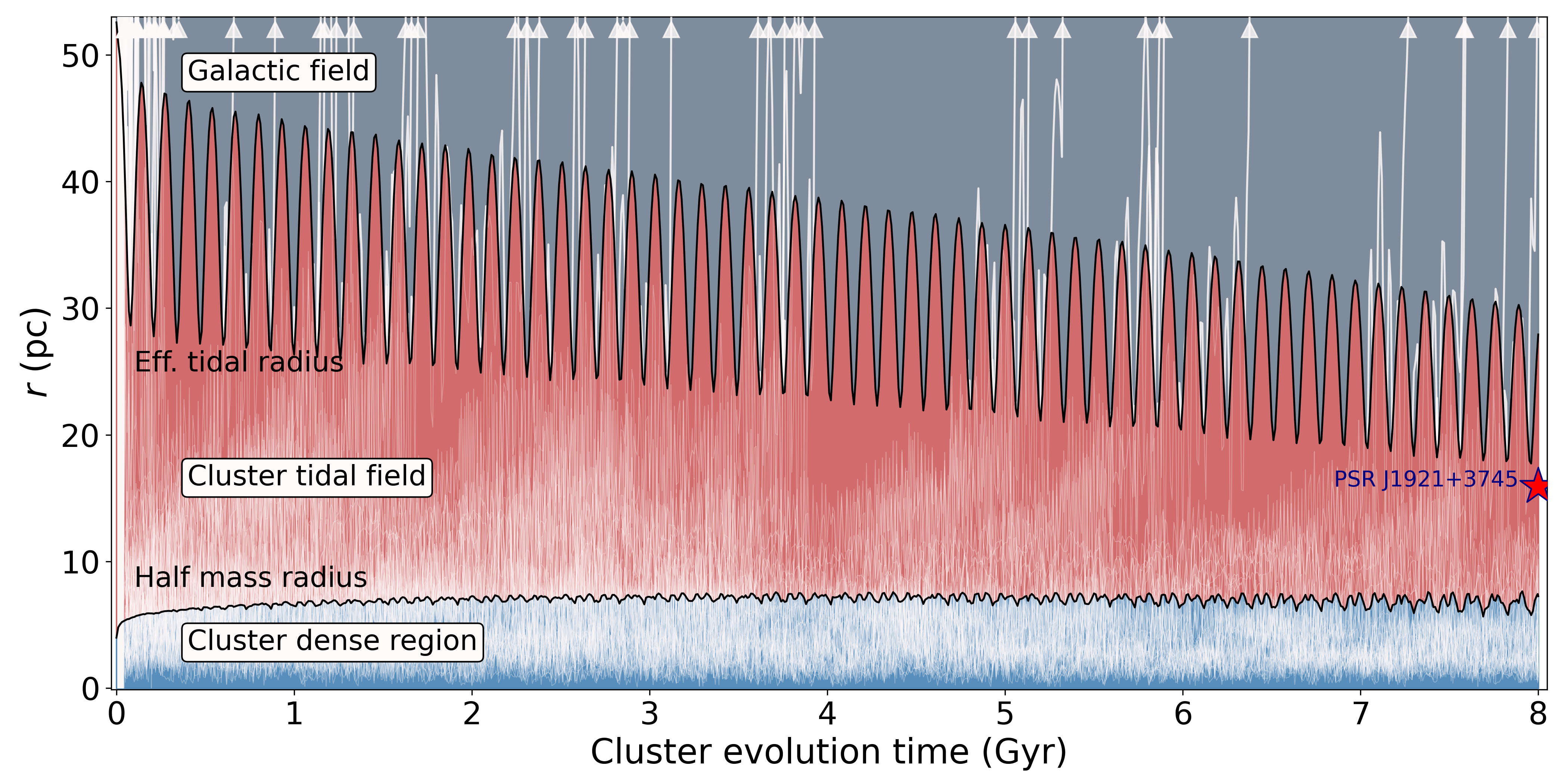}
    \caption{{\bf Radial evolution of NSs in an NGC 6791–like OC over 8 Gyr.} White curves trace each NS's distance from the cluster center. The lower and upper black curves show the half-mass radius and the time-varying effective tidal radius, respectively. Background shading marks three dynamical regimes: central dense core (blue), cluster tidal field (red), and Galactic field (grey). The red star indicates PSR~J1921$+$3745 at its present-day radius of $\sim$15.9 pc.}
\label{fig:ns_evo}
\end{figure*}

The radial evolution of these simulated NSs is shown in Fig.~\ref{fig:ns_evo}.
Owing to their large natal kicks, CCSN-formed NSs are almost entirely ejected within the first $\sim$40\,Myr. In contrast, ECSN-formed NSs with low kick velocities are retained much longer: 53 of the 96 escape over the subsequent gigayears through two-body relaxation and tidal stripping in the Milky Way potential, while the remaining 43 reside within the cluster’s effective tidal radius at the present epoch.
The effective tidal radius varies periodically as the cluster moves between Galactic pericenter and apocenter. The lower panel of Fig.~\ref{fig:ns_evo} shows the corresponding time evolution of the NS fractions in the dense region, tidal field, and escaped population, illustrating how low-kick ECSN NSs can gradually migrate outward from the cluster core into the tidal field before escaping.

Figure~\ref{fig:radec_2fig}b shows the normalized probability density of simulated NS positions at late times. The red contours trace the simulated stellar surface density at 8 Gyr and reveal the tidal tails, while the color map shows where simulated NSs are most likely to be found. The corresponding MeerKAT search area encloses an integrated simulated NS probability of $\simeq 7.4\%$, which is about 244 times higher than the random pulsar probability of  $\sim 0.03\%$.

These results indicate that NSs formed with low natal kicks can be retained in open clusters and later removed through tidal stripping, providing a natural explanation for the location of PSR~J1921$+$3745 in the cluster's tidal tail. 
The rarity of such detections is expected: only a small fraction of NSs are observable as radio pulsars due to beaming and finite radio lifetimes. PSR~J1921$+$3745 may therefore  represent the visible subset of a much larger, predominantly unseen NS population associated with the cluster (detailed in Supplementary material Section F).

%%%############### Conclusion ###############
%\section{Conclusion}\label{sec:conclusion}
The discovery and localization of PSR~J1921$+$3745 provides the first direct observational evidence of a NS in the tidal tail of an OC, offering a snapshot of a NS transitioning from a natal cluster environment to the Galactic field. Through tidal stripping, OCs can thus contribute to the Galactic disk's NS population, particularly those with low natal kicks. The presence of PSR~J1921$+$3745 in the tidal tail strongly supports a low-natal-kick formation scenario, consistent with ECSN and also compatible with other low-kick binary channels such as ultra-stripped supernovae or accretion-induced collapse (see detailed discussions in Supplementary Section F). Massive OCs can indeed be significant reservoirs of NSs, retaining them for Gyrs if they are born through low-kick channels. This contrasts with the typically rapid ejection expected for NSs from standard CCSNe of more massive stars and has profound implications for the overall NS budget of clusters. 

This work also potentially opens a new avenue for studying the co-evolution of NS populations and their host clusters, particularly the long-term dynamical interplay between NSs retained after low-kick supernovae and the cluster's tidal field. Observing NSs within tidal tails, as demonstrated here, allows us to directly probe the later stages of NS escape and the dissolution of clusters. Future systematic pulsar surveys targeting old and massive OCs, such as Berkeley 39 and NGC 188, are crucial. Detecting a  significant sample of NSs within OCs or their tidal features would help to 
(1)~provide constraints on the progenitor mass ranges for different supernova types (CCSN from massive stars vs. ECSN from massive-AGB stars with lower mass) and thus refine the stellar initial mass function;
(2)~allow for a more quantitative understanding of NS natal kick distributions, particularly the low kicks;
(3)~trace the dynamical history of cluster dissolution and the contribution of OCs to the Milky Way’s disk field NS population over time.

\section*{Conflict of interest}
The authors declare that they have no conflict of interest.

%\clearpage
\section*{Acknowledgments}
This work was supported by the National Natural Science Foundation of China (12588202, 12573040, 12533008, 12203100, 12103069, 12203045, 12273008, 12373109,\\ 12041305, 12173016, 1257030642, 12533003 and 12573030), the National Key R\&D Program of China (2023YFB4503305 and 2023YFE0110500), the Key R\&D Program of Shandong (2022CXGC020106), the Leading Innovation and Entrepreneurship Team of Zhejiang Province (2023R01008), the Key R\&D Program of Zhejiang (2024SSYS0012), the National SKA Program of China (2022SKA0130100 and 2022SKA0130104), the Natural Science and Technology Foundation of Guizhou Province ([2023]024), the Foundation of Guizhou Provincial Education Department (KY (2020) 003), the Natural Science Research Project of Anhui Educational Committee (2024AH050049), CAS Project (JZHKYPT-2021-06), and the Open Project Program of the Key Laboratory of FAST, Chinese Academy of Sciences.
We thank SARAO for approving the MeerKAT DDT request (DDT-20241205-LZ-01). TRAPUM observations used the FBFUSE and APSUSE computing clusters for data acquisition, storage, and analysis, funded by the Max-Planck-Institut f\"{u}r Radioastronomie and the Max Planck Society.
D.L. is a New Cornerstone Investigator. 
E.B. acknowledges support from the Max Planck Society. 
E.G. is supported by the Doctor Foundation of Qingdao Binhai University (BJZA2025025). 
We thank Matthew Bailes for helpful discussions.

\section*{Author contributions}
Lei Zhang, Xiaoting Fu, and Di Li led the manuscript preparation, with contributions and feedback from all co-authors. Xiaoting Fu was the principal investigator (PI) for the FAST observation, and Lei Zhang served as PI for the MeerKAT observation. Emma Carli coordinated the MeerKAT observing campaign, conducted the transient search and localization analysis, and processed the pulsar search data. Long Wang carried out the numerical simulations and dynamical modeling. Jiaqi Zhao and Craig O. Heinke analysed the X-ray data. Emanuele Dalessandro contributed to the optical dataset and archival counterpart search.
Yang Chen, Shi Dai, Jie Zhang, Meng Guo, Mengmeng Ni, Jiale Hu, Yi Feng, Pei Wang, and Qijun Zhi supported the proposal and execution of the FAST observations. Ewan D. Barr, Sarah Bunchner, and Alessandro Ridolfi proposed and coordinated the MeerKAT observations. Emma Carli, Long Wang, Jiaqi Zhao, Craig O. Heinke, Emanuele Dalessandro, Angela Bragaglia, Duncan R. Lorimer, Zhiyu Zhang, and Erbil G\"{u}gercino\u{g}lu made significant contributions to the interpretation of the results and refinement of the manuscript. All authors discussed the results and reviewed the manuscript.
\printcredits

\section*{Appendix A. Supplementary material}
Our processed data collection is publicly available via the Science Data Bank\footnote{\url{https://doi.org/10.57760/sciencedb.26295}}.
The FAST raw data related to this study are publicly available and can be obtained by sending a request to the FAST data centre. The MeerKAT raw data used in this study are available via the SARAO archive\footnote{\url{https://archive.sarao.ac.za}} under project ID 20241205-LZ-01. 
The {\it Chandra} X-ray data underlying this article are publicly available in the Chandra Data Archive\footnote{\url{https://cxc.harvard.edu/cda/}}.

\bibliographystyle{model3-num-names}
\bibliography{cas-refs}

\clearpage

\shorttitle{Zhang et al.}
\centerline{\bf\huge Supplementary Text}

\setcounter{section}{0}
\renewcommand{\thesection}{S\arabic{section}}
\setcounter{equation}{0}
\renewcommand{\theequation}{S\arabic{equation}}
\setcounter{figure}{0}
\renewcommand{\thefigure}{S\arabic{figure}}
\setcounter{table}{0}
\renewcommand{\thetable}{S\arabic{table}}

\appendix

\section{observations and Data reduction} \label{sec:obs}
\subsection{FAST data}
With the FAST 19-beam receiver at L-band (1000 – 1500\,MHz), we carried out a deep search of pulsars in NGC~6791 on 27 January 2023 through the observing project PT2022\_0075. We observed the cluster with an integration time of 1.83 hours, providing us with sufficient signal-to-noise ratio (SNR) for detecting faint pulsars. The central beam of the FAST 19-beam receiver was pointed at the nominal cluster center: $19^{\rm h}20^{\rm m}53^{\rm s}.0$, $+37^{\circ}46'18''.0$~\cite{Hunt2023}. 

The observation was cleaned from radio interference (RFI) and de-dispersed at several trial DMs, within a range of $\pm$100\,pc cm$^{-3}$ around the estimated DM from the Galactic electron density models (YMW16~\cite{Yao2017} = 50\,pc cm$^{-3}$ and NE2001~\cite{Cordes2002} = 75\,pc cm$^{-3}$), considering the source position and its inferred distance. For each de-dispersed time series, we performed a blind Fourier-domain periodicity search for fast-period pulsars, particularly in binary systems, using a \textsc{presto}~\cite{Ransom02}-based pipeline, with a Fourier-domain jerk search\cite{Andersen18} with a maximum linear Fourier drift rate $z_{\rm max} = 200$ and a maximum Fourier jerk drift rate $w_{\rm max} = 300$. We also split the 1.83-hour observations into 30 and 60 minute blocks so as to be sensitive to pulsar with orbital period as short as $\sim5$ hours\cite{Ng15}. We also performed a fast-folding search, which is particularly sensitive to slow-period pulsars, using a \textsc{riptide}-based~\cite{Morello20} pipeline. We set the minimum spin period for the search to 0.1\,s and the maximum to 10\,s. To ensure that long-period signals still have sufficient resolution at small duty cycles, we have appropriately adjusted the time resolution based on the size of the searched period. We also searched for possible single pulses by using a \textsc{heimdall}-based~\cite{Barsdell12} pipeline, with the aim to detect signals similar to Rotating Radio Transients (RRATs) and Fast Radio Bursts (FRBs). The trial DM range was up to 5,000\,pc cm$^{-3}$ and we matched the pulse width using a boxcar search up to 60\,ms.

\begin{figure*}
    \centering
    \includegraphics[height=5.8cm,width=15.5cm]{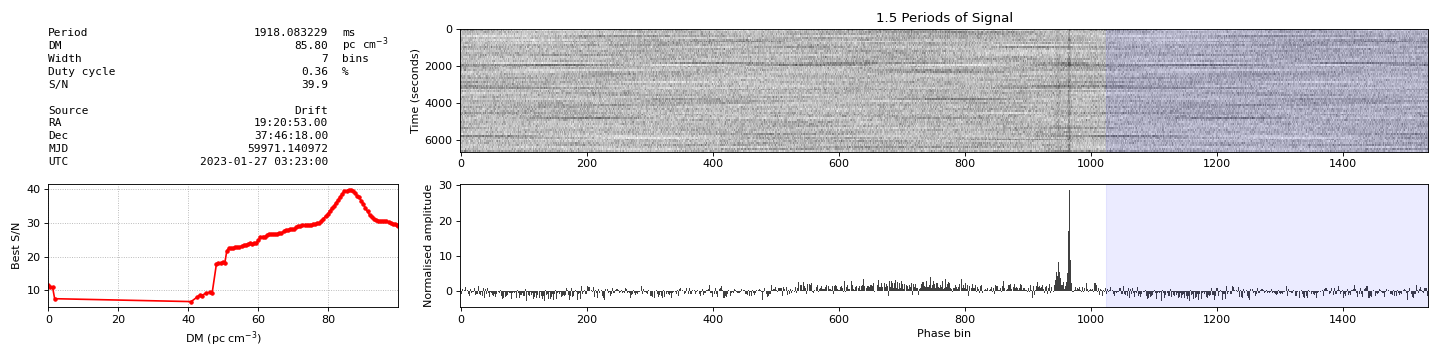}\\
    \includegraphics[height=11.2cm,width=7.9cm]{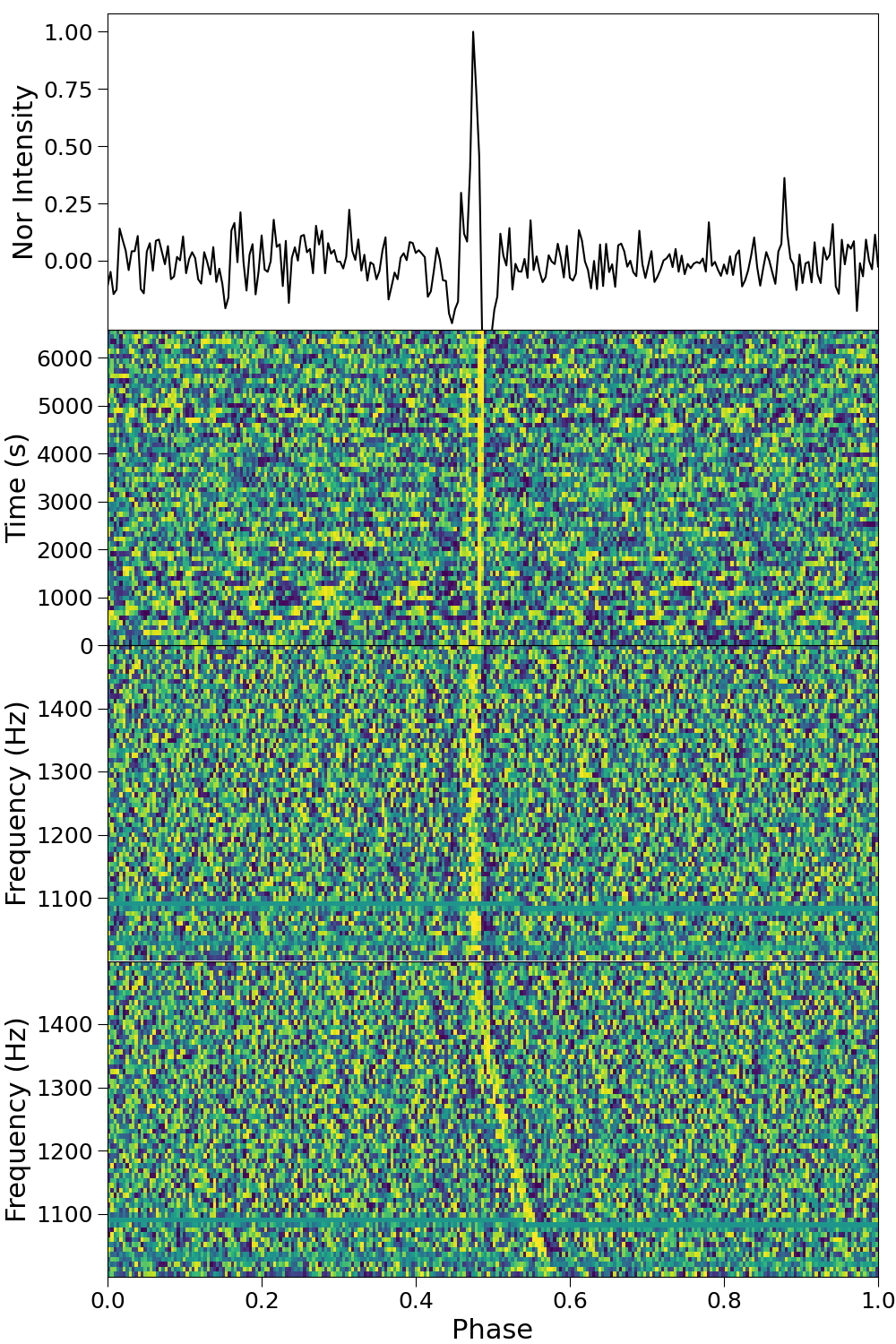}
    \caption{
    {\bf Discovery diagnostic plot of PSR~J1921+3745.}
     The pulsar has a spin period of $\sim$1.918\,s and a dispersion measure (DM) of 85.80\,pc\,cm$^{-3}$. The top panel shows 2.0\,hr of folded search-mode data recorded with the FAST 19-beam receiver on 2023 January 27. The signal was detected using a fast-folding algorithm implemented in \texttt{riptide}. The pulse profile (top-left) is folded with 1024 phase bins across one spin period; the blue shaded region spans an extended phase interval of 1.0 bin centered on phase 0.5 to ensure that at least one full pulse is visible. Each diagnostic panel (bottom) displays, from top to bottom: the integrated pulse profile, time-phase plot, and frequency-phase plot, shown both before and after de-dispersion.
     }
    \label{fig_FFA}
\end{figure*}

We discovered a 1.9-s pulsar signal at a DM of 85.8\,pc cm$^{-3}$ (hereafter rounded to 86 pc cm$^{-3}$ throughout the manuscript) with a signal-to-noise ratio (S/N) of 40 in a single beam (M08) using the Fast Folding Algorithm (FFA) implemented in \textsc{riptide} ( Fig.~\ref{fig_FFA}). 
To have independent confirmation of the new pulsar, we searched the archived data from FAST taken on 22 January 2022 using the \textsc{riptide}-based and \textsc{heimdall}-based pipeline. The signal was clearly detected in the observation (in two beams adjacent to the position of discovery beam M08), thereby providing confirmation of the discovery. In addition, three bursts were detected in the archived data (Fig.~\ref{fig_SPs-FAST}). With the pointing location based on three FAST detections, we obtained the optimal pulsar position to (Right Ascension, Declination) = ($19^{h}\,21^{m}\,59.26^{s}, +37^{\circ}\,46^{'}\,45.6^{''}$; J2000) with a 3\,arcmin error circle.

All the FAST observations were recorded with 8-bit sampling every 49\,${\mu}$s in pulsar search mode. In all the observations, the observing band from 1000\,MHz to 1500\,MHz was split into 4096 frequency channels and due to bandpass roll-off the effective band is from 1050\,MHz to 1450\,MHz.

\subsection{MeerKAT data}
Following the FAST detection of PSR~J1921+3745, a 1.98-hr MeerKAT observation was conducted on 17 December 2024 using director's discretionary time (DDT project DDT-20241205-LZ-01) at the position determined by FAST. The observation utilized the TRAPUM backend, which integrates the Filterbanking Beamformer User Supplied Equipment (FBFUSE) and the Accelerated Pulsar Search User Supplied Equipment 
(APSUSE)\footnote{\url{https://skaafrica.atlassian.net/wiki/spaces/ESDKB/pages/1591672833/User+Supplied+Equipment+USE\#FBFUSE/APSUSE}}.
The observation employed the UHF receiver~\cite{MKAT_UHF} with a central frequency of $f_{c} = 816$\,MHz and a bandwidth of $\Delta f = 544$\,MHz, divided into 2048 channels and sampled every 1.92\,ms. FBFUSE’s computing power synthesized 479 tied-array beams~\cite{Chen2021}, overlapping at a 75\% sensitivity level (at the beginning of the observation), to cover an approximately 5\,arcmin-radius circle using the full array during the pulsar's two-hour transit across the Observer's Meridian.
All TRAPUM beams, recorded as ``filterbank” search-mode files, were incoherently de-dispersed at the pulsar's DM of 86\,pc cm$^{-3}$. To reduce data volume, frequency channels were then summed in groups of 8, decreasing the channel count from 2048 to 256 (after cleaning with \textsc{iqrm}~\cite{iqrm}). After verification, the subbanded files were retained, and the original 2048-channel data were deleted.

\begin{figure*}
    \centering
    \includegraphics[width=0.95\linewidth]{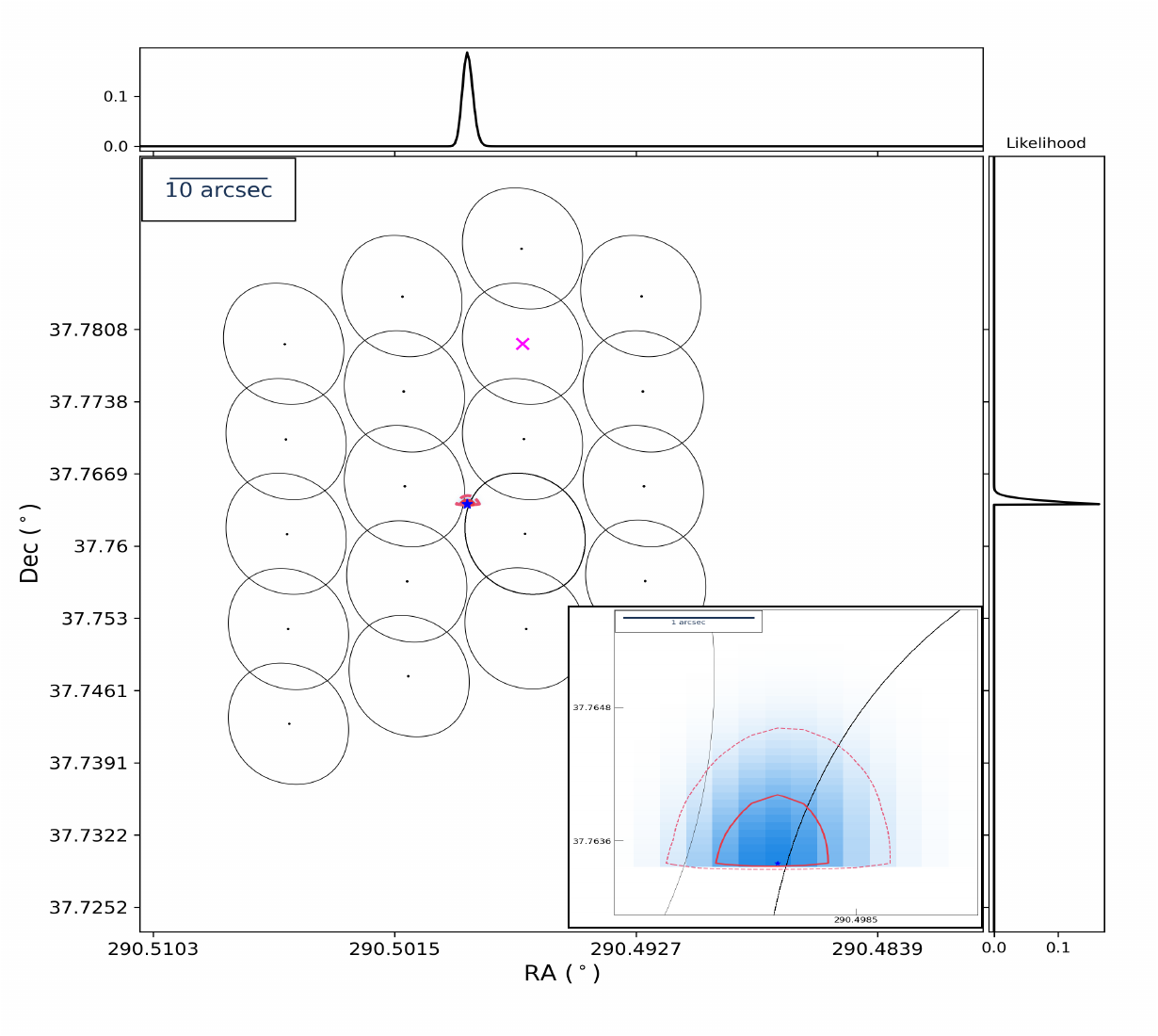}
    \caption{{\bf The localization of PSR~J1921+3745 using \texttt{seeKAT}.} The pulsar was detected in 19 MeerKAT tied-array beams, shown by the black outlines, which demarcate their 75\% sensitivity contour as determined by \textsc{mosaic} at the centre time and frequency of the observation. The inset plot on the bottom right provides a zoomed-in view of the maximum-likelihood position (see Table~\ref{tab:J1921_para}), which is indicated by a blue star. The 1$\sigma$ and 2$\sigma$ errors on the position are demarcated by the red continuous line and the red dashed line, respectively. The megenta cross marks the position of FAST beam M08. The closest MeerKAT beam to the localisation is 00033.}
    \label{fig_SeeKAT}
\end{figure*}

We first folded the data at the pulsar period of 1.918\,s using \texttt{psrfold\_fil} from the \textsc{pulsarx} suite~\cite{Men23}. The process included the removal of radio-frequency interference (RFI) with \textsc{pulsarx}'s \texttt{filtool} and masking of frequency channels significantly affected by RFI. PSR~J1921+3745 was initially detected in a coherent beam (00033). Subsequently, neighboring beams were folded with the discovery parameters to search for additional detections.To improve the pulsar localization, we used the \texttt{SeeKAT}\footnote{\url{https://github.com/BezuidenhoutMC/SeeKAT}} tied-array beam localization software~\cite{Bezuidenhout23}. \texttt{SeeKAT} performs a maximum-likelihood analysis by minimizing differences in signal-to-noise ratio (S/N) ratios between beam pairs, accounting for their positions and point spread functions (PSFs). The PSFs were calculated using the \texttt{Mosaic}\footnote{\url{https://wchenastro.github.io/mosaic_web/}} software~\cite{Chen2021}, based on the observation midpoint date, and central frequency. Weak beam detections that did not contribute significantly to the localization were excluded. Using 19 beam detections, we thus determined the pulsar's position as (Right Ascension, Declination) = ($19^{h}\,21^{m}\,59^{s}.70$, $+37^{\circ}\,45^{'}\,50^{''}.3$; J2000), with a 2$\sigma$ positional uncertainty of ($0.084^{s}$, $3.3^{''}$) (see Fig.~\ref{fig_SeeKAT}).

\begin{figure*}
    \centering
    \includegraphics[width=0.8\linewidth]{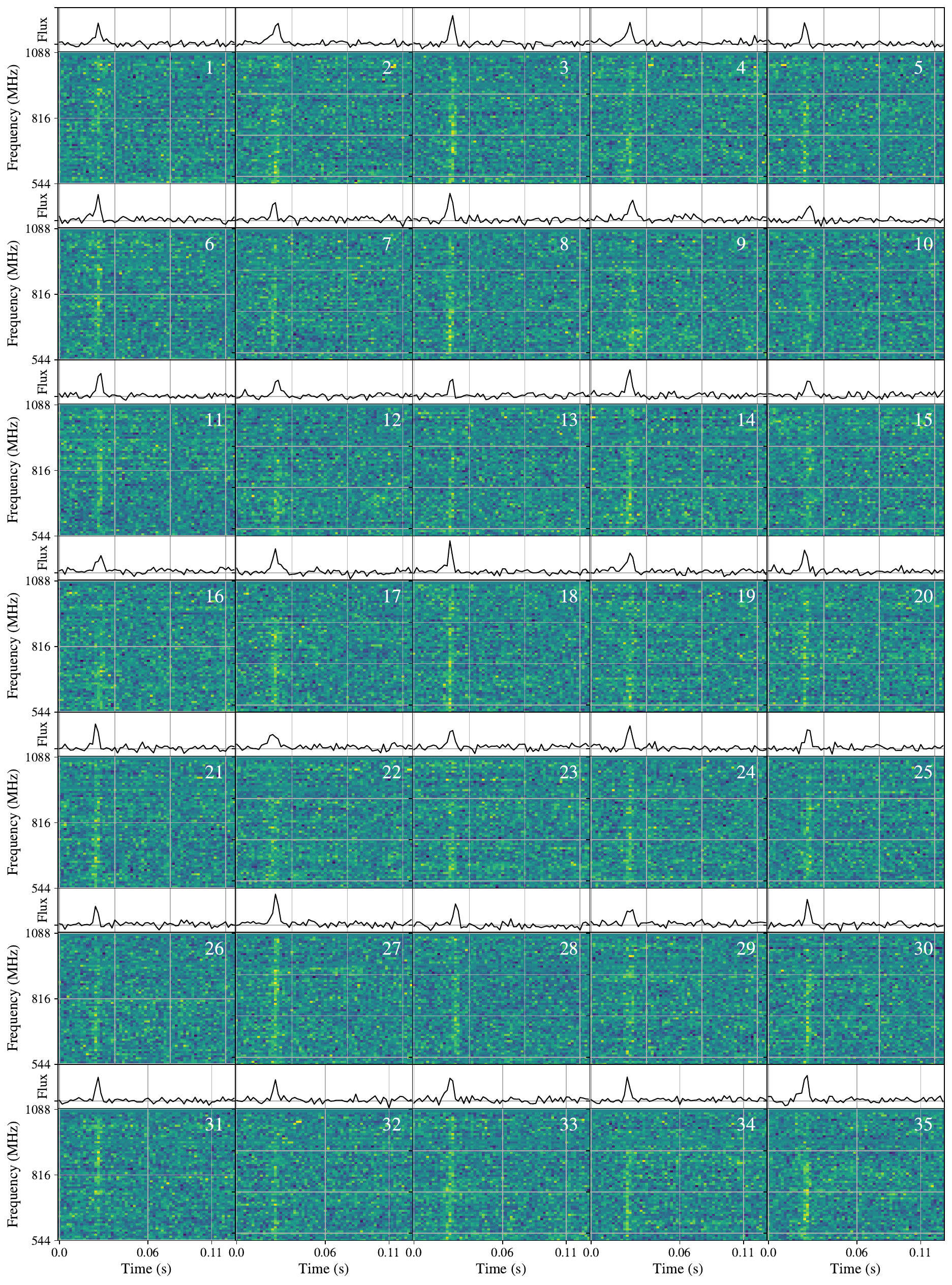}
    \caption{ {\bf Bursts from PSR~J1921+3745, shown as dynamic spectra and pulse profiles.} Top: frequency-integrated pulse normalized intensities that were detected in 2\,hour observation with MeerKAT data done on 2024-12-17 in beam 00033. Bottom: de-dispersed dynamic spectra of the pulses. The time samples are shown in raw resolution (1.9\,ms) and the UHF frequency band is downsampled to 64 channels.}     
    \label{fig_SPs-MKT}
\end{figure*}

We performed a comprehensive search of the entire observation for single pulses using \textsc{transientX}~\cite{Men24}, with the data pre-processed for RFI removal through a static channel mask and \texttt{pulsarX}'s filtool. The search was conducted with a maximum search width of 100\,ms across DMs ranging from 84 to 88\,pc cm$^{-3}$, and a minimum S/N threshold of 8. Approximately 100 candidates were identified from this observation, the majority of which originated from the incoherent beam, which is more prone to RFI due to its wider field of view. Following a detailed visual inspection of all single pulse candidates, 35 bursts were identified as not likely to originate from RFI or background noise fluctuations. These detections, displayed in Fig.~\ref{fig_SPs-MKT}, were all made in beam 00033. 

We also searched with \textsc{riptide} for long-period pulsars without significant acceleration from a binary companion, requiring a minimum FFA spectrum S/N ratio of 8, a period range of 0.2 to 10\,s and a DM range of 60 to 100\,pc cm$^{-3}$. However, no new pulsars were detected in any of the beams during this observing campaign.

\subsection{X-ray data}
\begin{figure*}
\centering
\includegraphics[width=12cm]{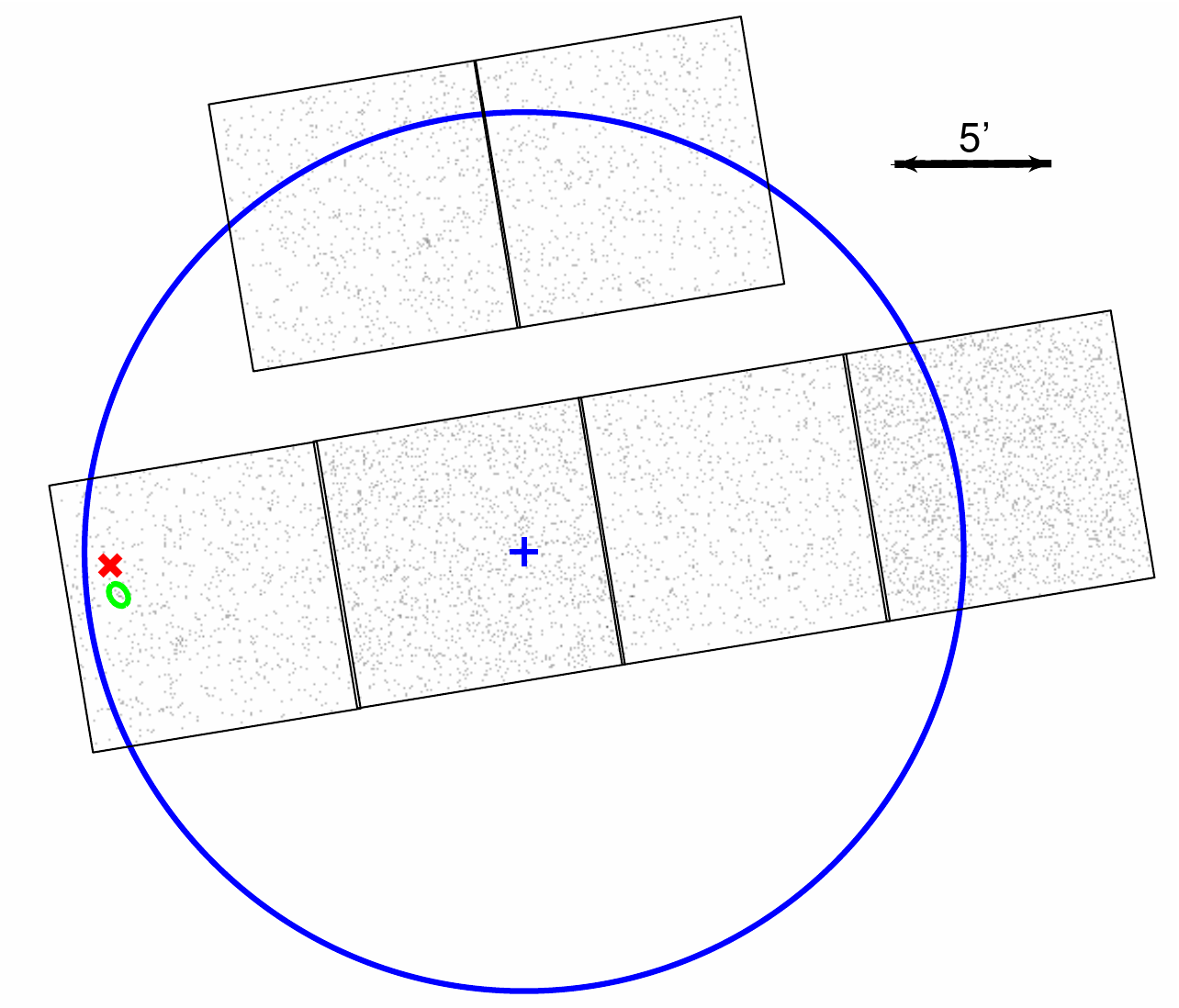}
\caption{ {\bf The \textit{Chandra} X-ray image of the OC NGC~6791 (Observation ID 4510) in the band 0.5--7 keV.} The black polygons show the field of view of this observation. The blue cross and circle show the center (R.A.=$19^{\rm h}20^{\rm m}53^{\rm s}.0$, Dec.=$+37^{\circ}46^{''}18^{'}.0$) and tidal radius (14~arcmin) of NGC~6791~\cite{Hunt2023}, respectively. The red ``X'' marks the radio position of PSR~J1921+3745, while the green ellipse indicates the nearest X-ray source to the pulsar, with an angular separation of 0.97 arcmin.
}
\label{chandra_fov}
\end{figure*}

To search for a potential X-ray counterpart to PSR~J1921+3745, we looked into the Chandra Source Catalog version 2.1 \cite{Evans24}. We first performed a cone search of 0.5 arcmin around the pulsar position but found no X-ray sources in the catalog. The nearest X-ray source, 2CXO~J192158.3+374454, is separated by 0.97~arcmin from PSR~J1921+3745 and is unlikely to be its X-ray counterpart. We then retrieved the only {\it Chandra} X-ray observation (Observation ID 4510) that covers the pulsar position for further investigation. However, the pulsar is located at the edge of {\it Chandra} S4 chip, with an off-axis angle of 13.8~arcmin (see Fig.~\ref{chandra_fov}). Consequently, the large point spread function might hide the X-ray emission from PSR~J1921+3745 within background emission over a large area, making its X-ray counterpart undetectable.  Nonetheless, we calculated the limiting energy flux within a 10-arcsec circle centered at the pulsar position, considering the point spread function. The estimated minimum energy flux required for a point source to be detected as a marginal source was found to be $1.2\times10^{-14}$~erg~cm$^{-2}$~s$^{-1}$ in the band 0.5--7 keV, which was considered as the estimate of the upper limit of the pulsar's X-ray flux. Future {\it Chandra} and XMM-Newton observations with the pulsar positioned near the optical axis are needed to identify its X-ray counterpart.

\section{Radio Flux Density Estimation}
We estimated the radio mean flux density of the pulsar using the radiometer equation~\cite{Lorimer04}:\\
\begin{equation}
S= \beta\frac{(S/N) t_{\text{sys}}}{G\sqrt{n_{p}t_{\text{int}}\Delta f}}\sqrt{\frac{W}{P-W}}, 
\end{equation}
where the sampling efficiency ($\beta$) is 1 for our FAST and MeerKAT 8-bit recording systems. The number of polarizations ($n_{p}$) is 2 for both FAST and MeerKAT observations. The pulse period ($P$) and the pulse width ($W_{\text{50}}$, $W_{\text{10}}$) of PSR~J1921+3745 are listed in Table~\ref{tab:J1921_para}. 
$T_{\text{sys}}$ is the system temperature, $G$ the telescope gain, $t_{\text{int}}$ the integration time and $\Delta f$ the observing bandwidth.

For the FAST discovery observation, the signal-to-noise ratio ($S/N$) was measured to be 44\footnote{Obtained using the \textsc{psrchive} tool \texttt{pdmp} applied to \textsc{dspsr}-folded observation data.\label{ft:pdmp}} with an integration time of $t_{\text{int}} = 6106$\,s. The effective bandwidth, $\Delta f$, is 350\,MHz, accounting for the masking of $\sim$12.5\% of channels affected by RFI, which were excluded from further processing. The system temperature, $t_{\text{sys}} \simeq 24$\,K\footnote{$t_{\text{sys}}$ is the sum of the receiver temperature and contributions from the sky, atmosphere, and ground spillover.\label{Tsys}}. Taking into account the position of beam M08, which reduces sensitivity by 8\% compared to the central beam M01, resulting in $G_{\text{M08}} = 14.8$\,K Jy$^{-1}$~\cite{Jiang20}, and the MeerKAT-determined pulsar position, which is 1.7~arcmin off the center of M08 (assuming a Gaussian beam sensitivity response), the sensitivity is further reduced by a factor of 2.5. This results in a gain of 5.92\,K Jy.

For the MeerKAT localization observation, the highest $S/N$ detection is equal to 20$^{\ref{ft:pdmp}}$ in beam 00033, achieved with an integration time of $t_{\text{int}} = 7142$\,s. The effective bandwidth is $\Delta f$ = 500\,MHz. Utilizing 60 antennas in the UHF band, the total telescope gain $G = 2.625$\,K Jy$^{-1}$, and the total system temperature $t_{\text{sys}} = 37.1$\,K. The latter includes contributions from the receiver temperature ($T_{\rm rec}$ = 20\,K)~\cite{Ridolfi22}, the sky ($T_{\rm sky}$ = 11.1\,K)~\cite{Zheng17, Price16}, atmosphere ($T_{\rm atm}$ = 1.5\,K)~\cite{Ridolfi22}, and ground spillover ($T_{\rm spill}$ = 4.5\,K)~\cite{Ridolfi22}. Using the observing and pulsar parameters described above, we apply the radiometer equation to estimate the mean flux density of the pulsar. The results are $S_{\text{FAST}} = 6.7 \pm 1.1$\,$\mu$Jy in the FAST L band and $S_{\text{MeerKAT}} = 8.1 \pm 1.4$\,$\mu$Jy in the MeerKAT UHF band. The uncertainty stated arises from the uncertainty in the measured pulse width. 

\begin{figure*}
    \centering
    \includegraphics[width=0.9\linewidth]{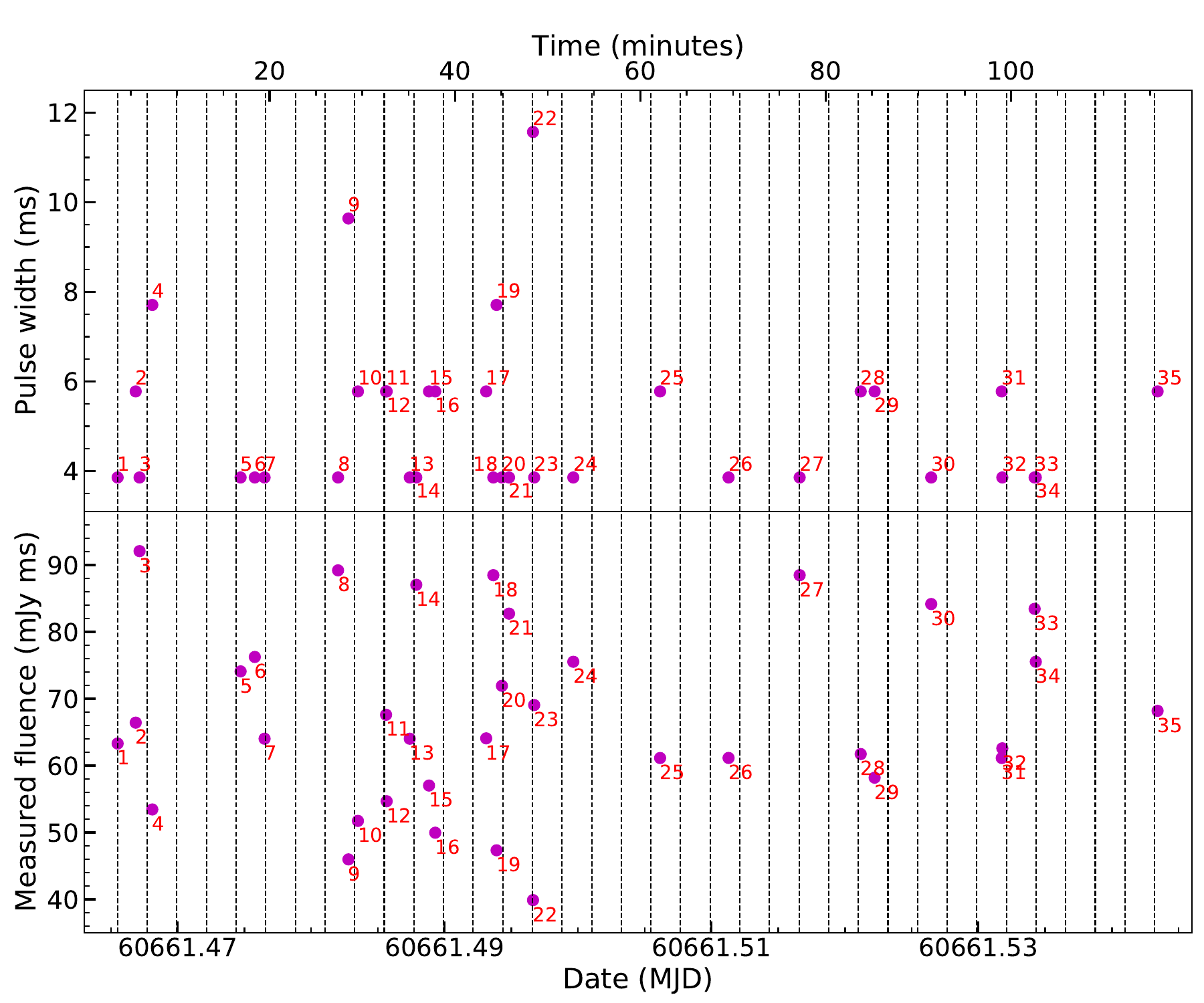}
    \caption{ {\bf Bursts from PSR~J1921+3745, showing pulse width and measured fluence.} Top: Pulse width of pulses detected during the 2-hour MeerKAT observation on 2024-12-17. Bottom: Measured fluence of the pulses.}      
    \label{fig_MKTSPs_wf}
\end{figure*}

We used the values above in the following equation to calculate the single pulse flux density~\cite{Cordes03}:
\begin{equation}
S_{\text{pulse}}=\beta \frac{(S/N)t_{\text{sys}}}{G\sqrt{n_{\text{p}}w_{\text{broadened}}\Delta f}},
\end{equation}
where $w_{\text{broadened}}$ is the observed width of the burst.  The properties of the pulses detected with MeerKAT, as reported by \textsc{transientx}, in Table~\ref{tab:MKT-SPs} and Fig.~\ref{fig_MKTSPs_wf}.

\section{Distance uncertainties}
\begin{figure*}
    \centering
    \includegraphics[width=0.7\linewidth]{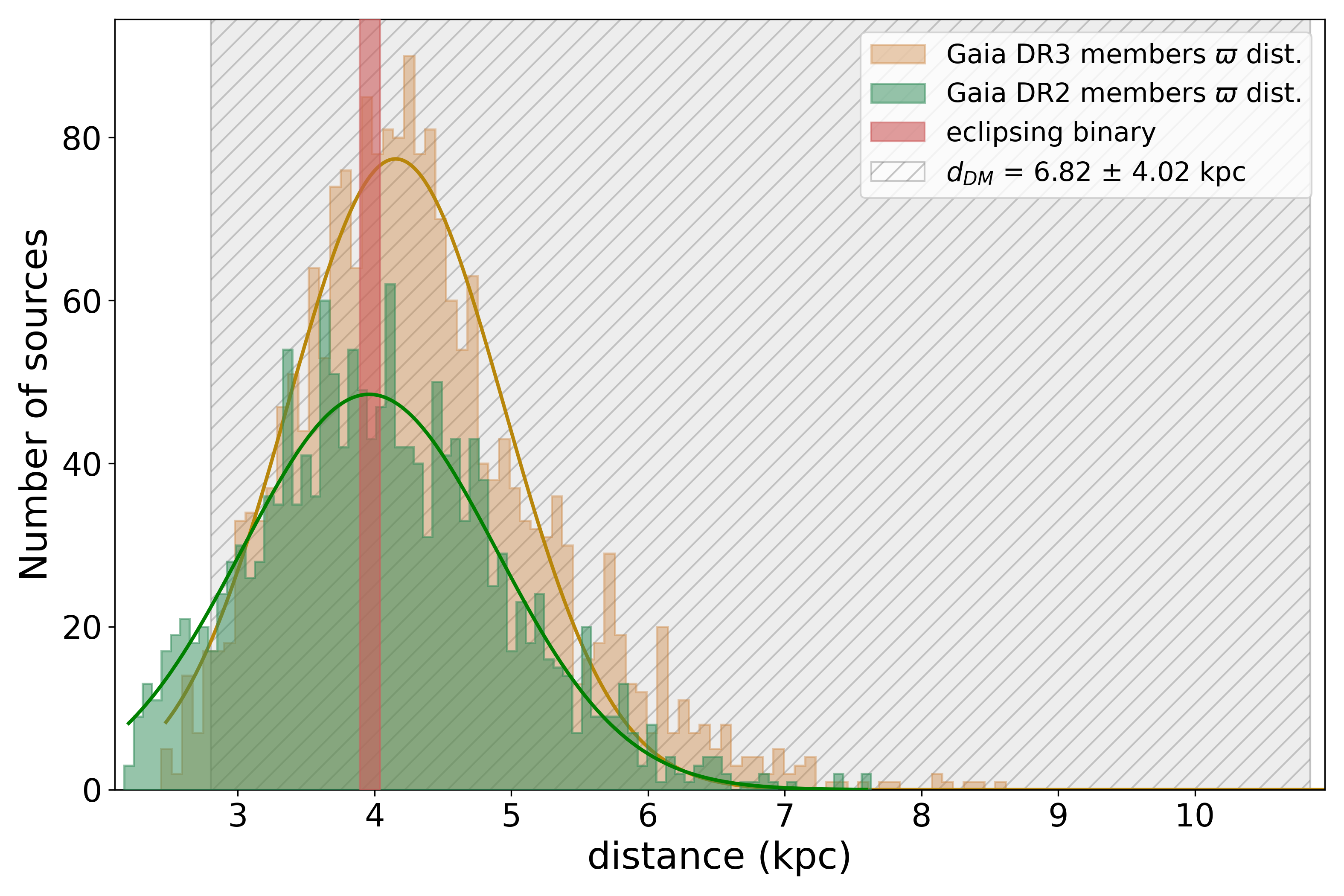}
    \caption{NGC 6791 member star distances based on Gaia DR2 \cite{CG2020} and Gaia DR3 \cite{Hunt2024} with parallax ($\varpi$) corrections \cite{BJ2018, BJ2021}. The distance range of NGC 6791 eclipsing binary is marked with the vertical red line. The distance range of PSR~J1921$+$3745 derived from DM  is  illustrated with a grey region.}
\label{gaia_dist}
\end{figure*}

\textbf{Distance from Gaia parallax.}
To compare the pulsar with individual cluster member stars' distance from the Gaia parallax, we use the Bailer-Jones distance catalog\cite{BJ2018, BJ2021}, which provides distance estimates from Gaia DR2 and (e)DR3 parallaxes, accounting for measurement biases. As shown in  Fig.~\ref{gaia_dist}, these member stars form a distribution centered around $\sim$4 kpc,  with a standard deviation of $\sim$0.9 kpc. The breadth of this distribution is dominated by the individual parallax uncertainties \cite{Xu2025}. Although the average parallax error of member stars is less than $\sim$0.1 mas, it leads to large distance uncertainties at this distance.

\textbf{DM distance of the pulsar.}
The DM of PSR~J1921$+$3745 from our MeerKAT observation yields d$_{\rm NE2001}\sim$4.84 kpc, d$_{\rm YMW16}\sim$9.07 kpc, and d$_{\rm NE2025}\sim$6.56 kpc, with the three most commonly used electron density models (see Table \ref{tab:J1921_para})
The mean value of the distance is d$_{\rm mean}$ = (d$_{\rm YMW16}$+d$_{\rm NE2001}$+d$_{\rm NE2025}$)/3 = 6.82 kpc. Their standard deviation serves as the systematic spread range: $\sigma_{sys}$ = std(d$_{\rm YMW16}$, d$_{\rm NE2001}$, d$_{\rm NE2025}$) = 2.13 kpc.
We assign a statistical uncertainty of 50 \% to each model following suggestions in the literature \cite{Deller2019} and get the statistical spread $\sigma_{stat}$ = 0.5 d$_{\rm mean}$ = 3.41 kpc. 
Adding these in quadrature gives the total 1 $\sigma$ error: $\sigma_{tot} = \sqrt{\sigma_{sys}^2 + \sigma_{stat}^2} =4.02$ kpc. We therefore adopt d$_{\rm DM}$ = 6.82 $\pm$ 4.02 kpc.

\section{Pulsar Number Estimation along a Line of Sight}
To estimate the number of detectable pulsars along the line of sight to PSR~J1921$+$3745, we adopt the volumetric Galactic pulsar distribution model described by Verbiest et al.~\cite{Verbiest2012}, which parameterizes the spatial density of normal (non-millisecond) pulsars as a function of Galactocentric radius and height above the Galactic plane:
\begin{equation}
 \rho (R,z)=\frac{N}{V} \simeq R^{B}~\exp\left[-\frac{|z|}{E} - C\frac{R - R_0}{R_0}\right]~\text{kpc}^{-3},
\end{equation}
where $N$ being the number of pulsars per volume $V$ and constants $R_0 = 8.5$\,kpc is the Galactocentric radius of the Sun, $B = 1.9$, $C = 5$, and $E = 330$\,pc is the exponential scale height for normal pulsars~\cite{Lorimer2006}. $R$ is the Galactocentric radius and $z$ is the height above/below the Galactic plane.

This density profile yields the following line-of-sight distribution in Earth-centered Galactic coordinates (distance $D$, Galactic latitude $b$, and longitude $l$)~\cite{Verbiest2012}:
\begin{equation}
p(D|l,b) \propto R(D)^{1.9}~\exp\left[-\frac{|z(D)|}{E} - 5\frac{R(D) - R_0}{R_0}\right]D^2
\end{equation}
with
\begin{equation}
z(D,b) = D \sin b
\end{equation}
\begin{equation}
R(D,l,b) = \sqrt{R_0^2 + (D \cos b)^2 - 2R_0 D \cos b \cos l}.
\end{equation}
We restrict our estimate to the population of slowly rotating pulsars ($P > 1.5$\,s), which comprise approximately 10\% of the normal pulsar population beaming toward Earth~\cite{Lorimer2006}. Integrating the modeled volumetric density along the line of sight from $D = 0$ to $4.2$\,kpc—the distance to NGC 6791—at Galactic coordinates $(l = 69.96^\circ, b = 10.90^\circ)$, we estimate a surface density of detectable pulsars above a 1.4\,GHz luminosity threshold of 0.1\,mJy kpc$^2$ to be $\sim$0.0096 deg$^{-2}$. Over the $\sim$0.6 deg$^2$ area encompassing the cluster and its tidal tails, this implies a chance alignment probability of $\sim$0.6\%.

To assess the robustness of this result, we re-evaluated the expected surface density using the volumetric pulsar distribution model from Verbiest et al.~\cite{Verbiest2012}, performing 100 Monte Carlo realizations that varied the vertical scale height ($h_z = 330 \pm 50$\,pc), the radial power-law index ($R^{1.9 \pm 0.2}$), and the overall normalization (±10\%). Extending the integration to 100 kpc, we find an average surface density of $0.02 \pm 0.01$ long-period pulsars per deg$^2$, corresponding to a revised chance-alignment probability of $2 \pm 1$\%. This value is only weakly sensitive to the assumed integration depth, as the pulsar density along this high-latitude sightline remains low beyond $\sim$4\,kpc. While this probability alone does not definitively establish association, its low value—combined with the pulsar’s precise spatial alignment with the cluster’s tidal structure and consistent DM-based distance—strongly supports a physical connection between PSR~J1921$+$3745 and NGC 6791.

\section{Cluster association}
\begin{figure*}
\centering
\includegraphics[height=12cm,width=12cm]{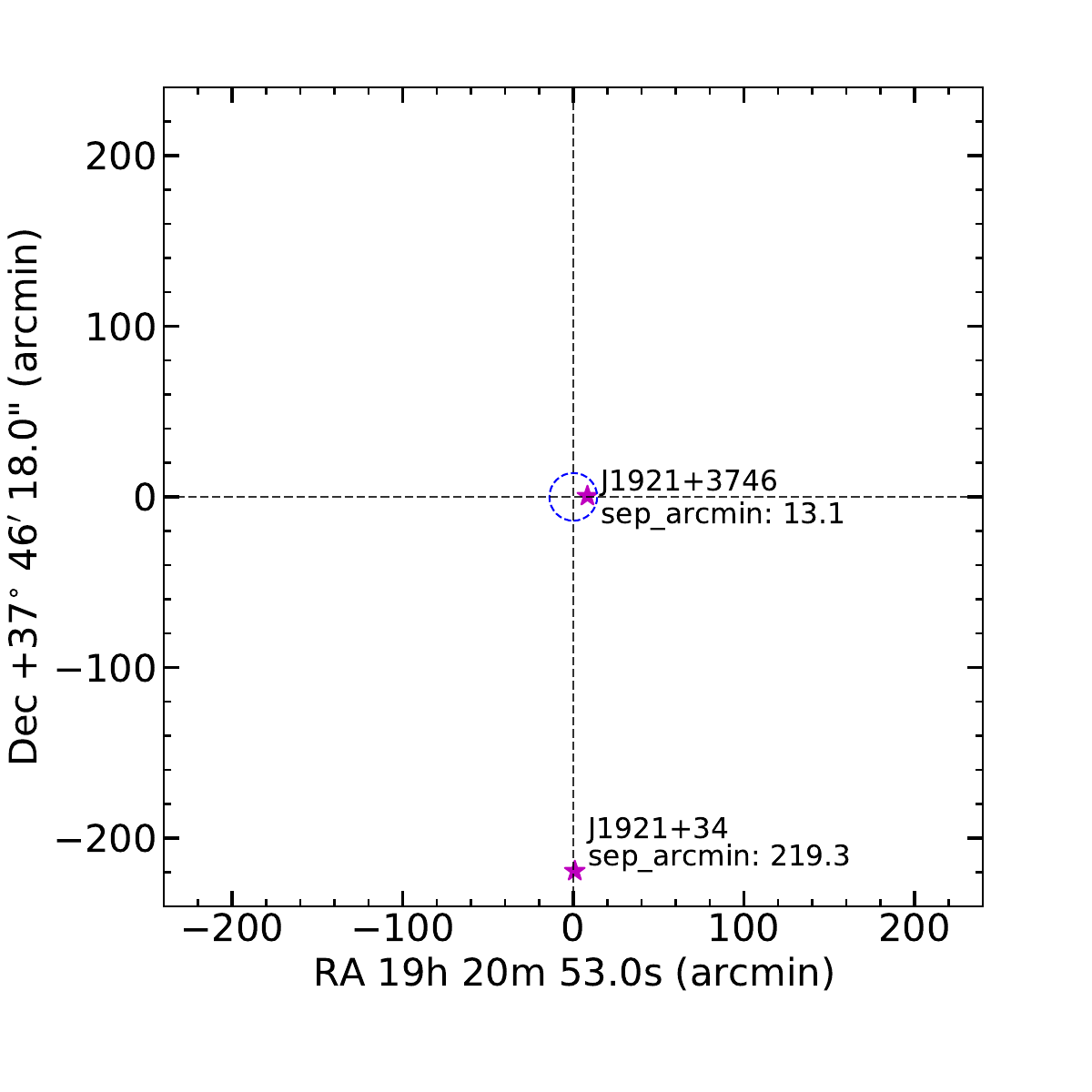}
\caption{{\bf Positions of PSR~J1921+3745 and its nearest known pulsar J1921+34}. The concentric blue dashed circle with radii of 14$^{'}$ shows the radius, centered at $19^{\rm h}20^{\rm m}53^{\rm s}.0$, $+37^{\circ}46^{''}18^{'}.0$.}
\label{fig_nearbyPSRS}
\end{figure*}

To assess the potential for source confusion and to search for any persistent emission from PSR~J1921$+$3745, we inspected archival radio data. 
The nearest catalogued pulsar in the ATNF Pulsar Catalogue\footnote{\url{https://www.atnf.csiro.au/research/pulsar/psrcat/}\label{ATNF}} is located more than 200 arcmin from PSR~J1921$+$3745 (see Fig.~\ref{fig_nearbyPSRS}), precluding source confusion. No persistent radio source was detected at the pulsar’s precise MeerKAT position in either the Rapid ASKAP Continuum Survey (RACS, central frequency 1367.5 MHz; \cite{Duchesne2024}) or the NRAO VLA Sky Survey (NVSS, central frequency 1400 MHz; \cite{Condon1998}). This non-detection of steady emission is consistent with the faint, pulsed nature of PSR~J1921$+$3745. 

We search for counterparts to PSR~J1921$+$3745 at other wavelengths. No associated supernova remnant (SNR) or pulsar wind nebula (PWN) was found in existing radio or X-ray surveys. No X-ray point source is detected within a 0.5 arcmin radius of the pulsar's MeerKAT position in the Chandra Source Catalog (version 2.1)\cite{Evans24}, indicating no ongoing accretion.
No optical counterpart is identified within a 3 arcsec search radius in Gaia DR3 \cite{Gaia2023}, or in the Hubble Space Telescope archival data. The MegaCAM/CFHT photometry used to identify the tidal tails of NGC 6791\cite{Dalessandro2015}, being the deepest photometry currently available, also reveals no optical source at the pulsar's position. The lack of a X-ray counterpart strongly constrains the accretion rate onto the pulsar, ruling out a system with significant ongoing mass transfer. Furthermore, we use the non-detection in deep MegaCAM/CFHT optical imaging to constrain the nature of any putative companion. By comparing the PARSEC stellar isochrone \cite{Bressan2012} with the cluster's age, distance, and metallicity \cite{Fu2022},  our limiting magnitude of g$\approx$22.5 mag corresponds to M$\sim$0.7 M$_{\odot}$. Any main sequence companion, if it exists, must have a mass M$\lesssim$0.7 M$_{\odot}$. 

To evaluate the robustness of PSR J1921+3745's association with NGC 6791, we compared its DM-derived distance with the cluster distance inferred from three independent techniques: \textit{i)} Gaia‐based isochrone fitting, \textit{ii)} Gaia parallaxes of individual member stars, and \textit{iii)} eclipsing binary analysis. Recent fits to Gaia DR2 and DR3 photometry place the cluster center at 4.189 kpc \cite{CG2020} and 4.231 kpc \cite{Hunt2024}, respectively. These estimates are dominated by systematic uncertainties tied to reddening, metallicity and the adopted stellar models. Propagating the individual Gaia parallaxes of confirmed members yields a mean distance consistent with the isochrone value but a one-sigma line-of-sight spread of $\sim$0.9 kpc (see Supplementary material
Section C) because of parallax-to-distance inversion effects \cite{Xu2025}. Analysis of the cluster’s eclipsing binaries, possibly the most accurate method, gives the distance of $\sim$ 4.87 - 5.13 kpc before the extinction correction and $\sim$3.89 - 4.04 kpc after \cite{Brogaard2011}.

The DM-derived distance of a pulsar relies on the electron density model with substantial uncertainties \cite{Deller2019}. Adopting the  three most widely used Galactic electron-density models NE2001  \cite{Cordes2002}, YMW16 \cite{Yao2017}, and NE2005 \cite{Ocker2026}
we obtain the distance of d$_{\rm NE2001}\sim$4.84 kpc, d$_{\rm YMW16}\sim$9.07 kpc, and d$_{\rm NE2025}\sim$6.56 kpc for PSR~J1921$+$3745, respectively (see Table~S1). Following the $PSR\pi$ comparison to VLBI parallax \cite{Deller2019}, we combine a 50\% statistical allowance with the model-to-model spread and obtain d$_{\rm DM}$ = 6.82 $\pm$ 4.02 kpc (see Supplementary material Section C for calculation details). Within 1 $\sigma$ this range overlaps the Gaia and binary based distances, supporting an association between PSR J1921+3745 and NGC 6791.

To further strengthen the association, we evaluated the probability of a chance alignment with an unrelated field neutron star. 
The recently reported timing solution for PSR~J1921$+$3745 provides a measurement of the pulsar spin-down rate \cite{LiuXJ2026}. The inferred characteristic age of $\sim$7.8 Myr and surface dipole magnetic field strength of $\sim2.8\times10^{12}$ G place the pulsar within the population of canonical isolated slow pulsars on the $P$–$\dot{P}$ diagram (Fig.~\ref{fig_J1921}b). 
We note that characteristic ages inferred from spin-down are model-dependent and do not always provide an accurate measure of the true age of a neutron star (e.g. \cite{Zhang2026}). Nevertheless, the timing properties clearly indicate that PSR~J1921$+$3745 is an ordinary radio pulsar rather than a high-magnetic-field neutron star or magnetar. 
An ``interrupted-recycling'' explanation for the long spin period is also unlikely, since even globular clusters are generally not dense enough to halt mass transfer sufficiently rapidly to produce the observed $P$–$\dot{P}$ combination. 
The updated timing solution does not alter the dynamical interpretation presented in this work. Rather, it places PSR~J1921+3745 within the population of canonical slow pulsars on the $P$–$\dot{P}$ diagram. While the positional coincidence and dynamical evidence support an association with NGC~6791, the evolutionary pathway leading to its present-day radio-emitting state remains uncertain.

The recently reported timing solution for PSR~J1921$+$3745 disfavors an interpretation as a newly formed young neutron star or magnetar \citep{LiuXJ2026}. We nevertheless searched for nearby star-forming regions and young stellar associations around the pulsar position. Our inspection of the WISE Catalog of Galactic H\,II Regions V2.3 \cite{Anderson2014} reveals no nearby star-forming regions, while searches of the open cluster catalogues \cite{Hunt2023, Hunt2024} identify no young star clusters or OB associations in the vicinity of the pulsar. These results further support the interpretation that PSR~J1921$+$3745 is an evolved neutron star associated with NGC~6791.

The likelihood of a chance alignment, as a generic field pulsar, is only $\sim$0.6\% over the entire cluster area including the tidal tails, and is $\sim$0.03\% for the MeerKAT 6 arcmin observing area. For comparison, according to our N-body simulation, a NS probability in the MeerKAT observing area is $\sim$ 244 times higher.
Using the volumetric spatial distribution model for normal pulsars from Verbiest et al.\cite{Verbiest2012}, we integrated the expected number of detectable sources along the line of sight toward NGC~6791 ($l = 69.96^\circ$, $b = 10.90^\circ$), out to its distance of 4.2\,kpc. Applying a spin period threshold of $P > 1.5$\,s and a luminosity cutoff of 0.1\,mJy\,kpc$^2$—below the estimated luminosity of PSR~J1921$+$3745 (0.16\,mJy\,kpc$^2$)—we estimate a surface density of only $\sim$0.0096 pulsars per square degree in this direction. Over the 0.6\,deg$^2$ area encompassing the cluster and its tidal structures, this yields a chance alignment probability of merely $\sim$0.6\% (see Supplementary material Section D for details). In the same simulation, if we integrate out to 100 kpc, the chance alignment probability increase to $2\pm1\%$. This is a gross overestimate as the current pulsar population is largely based on the sources in the disk, where most of pulsar searches were conducted and most of the currently known pulsars are located. The derived low probability thus strongly supports the conclusion that PSR~J1921$+$3745 is not a random interloper but is physically associated with the stellar population of NGC~6791.

A direct proper-motion measurement would provide an important independent constraint on the association between PSR J1921+3745 and NGC 6791. However, the pulsar is extremely faint, with a mean flux density of only several $\mu$Jy, making precision VLBI astrometry currently impractical. The presently available timing baseline is also insufficient to derive a reliable timing proper motion. 
In practice, obtaining a meaningful timing proper-motion measurement for such a faint long-period pulsar is likely to require several additional years of timing observations.
Continued long-term timing observations with FAST may eventually place useful constraints on the pulsar’s transverse motion.

\section{N-body simulations}
To explore the formation scenario of PSR~J1921$+$3745, we perform $N$-body simulations of its host cluster NGC 6791 using the \petar~ code \cite{petar}. This code accurately models star cluster dynamics with the \textsc{fdps} framework \cite{Iwasawa2016} and the Hermite method with slow-down algorithmic regularization \textsc{sdar} \cite{Wang2020a}. It also tracks single and binary stellar evolution via the \textsc{sse/bse} code \cite{Hurley2000,Hurley2002,Banerjee2020}, and models Galactic tidal forces using \textsc{galpy} with the MWPotential2014 model \cite{Bovy2015}.

We assume NGC 6791 formed 8 Gyr ago with an initial mass of $10^5 M_\odot$ and a half-mass radius of 4 pc. The simulation begins after gas expulsion in a virialized, gas-free state, with all stars sharing the same formation time and metallicity ($Z\approx0.03$ \cite{Bragaglia12, Fu2022}). The positions of stars follow a Plummer sphere distribution \cite{Plummer1911}, the Kroupa IMF is used \cite{Kroupa2001}, and there is no primordial mass segregation or primordial binaries. 

\begin{figure*}
    \centering
    \includegraphics[width=0.7\textwidth]{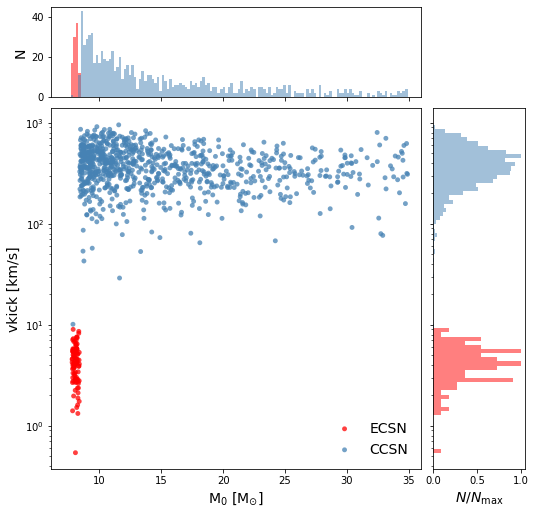}
    \caption{NSs Kick velocity distribution as a function of initial stellar mass in our simulated cluster. The ECSN- and CCSN-formed NSs are marked with red and blue, respectively. The histogram of the kick velocity is normalized to the max number of the corresponding subset.}
    \label{fig:vkick}
\end{figure*}

Upon formation, NSs receive natal kicks arising from asymmetries in supernova explosions. If this kick velocity exceeds the cluster’s escape velocity, the NS is promptly ejected. Observations indicate that these kick velocities typically follow a Maxwellian distribution with a 1D dispersion of 265 km/s \cite{Hobbs2005}, causing most NSs to escape their clusters soon after formation. However, NSs formed via ECSNe, which originate from lower-mass progenitor stars, are expected to receive significantly lower kicks (dispersion $\sim 3$ km/s), a scenario likely applicable to PSR~J1921$+$3745. 
Fig.~\ref{fig:vkick} illustrates the kick velocity distributions for the NS populations in our simulation, distinguishing between those formed via CCSNe (median kick $\approx$ 369 km/s) and ECSNe (median kick $\approx$ 4.2 km/s). 

In  the \textsc{petar} simulations, ECSNe are enabled when  two core-mass conditions are satisfied: the core mass at the onset of the AGB lies in the range $1.6 \leq M_{\rm c,BAGB}/M_\odot \leq 2.25$, and the core subsequently grows to $M_{\rm c} \geq 1.372\,M_\odot$. Here $M_{\rm c,BAGB}$ denotes the He-exhausted core mass at the onset of the AGB phase. Once these conditions are met, ECSN is triggered.
The threshold $M_{\rm c}=1.372\,M_\odot$ is the electron-capture collapse threshold adopted in the SSE/BSE prescription. It represents the mass at which an ONe core becomes unstable to electron captures, close to but slightly below the canonical Chandrasekhar mass \cite{Nomoto1987, Fryer2012}.

We note that the ECSN channel in the fiducial N-body simulation is implemented through the stellar evolution prescription adopted in SSE/BSE, while the initial cluster model does not include primordial binaries. Thus, the simulation should not be interpreted as distinguishing whether PSR~J1921$+$3745 originated from a single-star or binary ECSN progenitor. 

This distinction does not affect the main dynamical conclusion of this work: the retention and later tidal stripping of an NS in an open cluster are primarily controlled by the natal kick and subsequent cluster dynamics. Recent reviews emphasize that ECSNe can occur through both single-star and binary channels, and that both are associated with relatively small ejecta masses and low natal kicks compared with canonical CCSNe \cite{Wang2026}. Therefore, a binary-origin ECSN would remain fully consistent with our proposed low-kick formation scenario.

ECSNe are not the only possible route to forming low-velocity NSs. Ultra-stripped supernovae in close binaries can eject very little mass and are expected to produce relatively small natal kicks  \citep{Tauris2015} as well. Accretion-induced collapse (AIC) of ONe white dwarfs in interacting binaries can also form NSs through a low-ejecta-mass channel \citep{Tauris2013}. If such binaries are disrupted during or after the explosion/collapse, they may leave behind isolated NSs with low systemic velocities.

These channels are therefore dynamically compatible with the scenario discussed in this work. The key requirement for retaining an NS in a star cluster for several gigayears is not the detailed progenitor label, but the low post-formation velocity of the NS or binary center of mass. Our simulations explicitly model ECSN-formed NSs as a representative low-kick channel. A quantitative comparison between ECSNe, ultra-stripped SNe, and AIC would require a full binary-population synthesis coupled to cluster dynamics, which is beyond the scope of the present work. We therefore interpret PSR~J1921$+$3745 as evidence for a low-kick NS formation channel, with ECSN being the channel explicitly followed in our simulations.

To reproduce the cluster’s tidal tail, we adopt the method described by \cite{Wang2021}. We initialized the cluster's orbit using NGC 6791’s present-day position and velocity (Table~\ref{tab:init}), inverting the velocity vector to integrate the orbit backward by 8 Gyr within the Galactic potential to determine its birth position. We then re-inverted the velocity to serve as the initial condition for our simulation, evolving the cluster forward from birth to the present epoch. This approach allows us to self-consistently trace the cluster's dynamical evolution, including its interaction with the Galactic potential and the subsequent formation of tidal streams. As tidal tails develop, the cluster morphology inevitably deviates from spherical symmetry. To characterize the dynamical structure of the cluster throughout this long-term evolution, we use the effective tidal radius, calculated under the point-mass approximation.

Along the cluster's eccentric Galactic orbit, the strength of the external tidal field varies with Galactocentric distance. Near pericenter, the Galactic tidal field is stronger, leading to a smaller effective tidal radius; near apocenter, the tidal field is weaker and the effective tidal radius increases. The effective tidal radius therefore varies quasi-periodically over the orbital cycle, in addition to any long-term decrease driven by cluster mass loss.

Our simulations help explain why previous pulsar searches in OCs have been unfruitful in identifying resident NSs. Firstly, a significant fraction of OCs are born with relatively low initial masses (e.g., less than several 10$^3 M_\odot$ \cite{Hunt2024, Almeida2025}). Such clusters are dynamically fragile and tend to dissolve into the Galactic field over a short timescale \cite{Baumgardt2003, Almeida2025}. Any NS formed within these less massive, shorter-lived clusters would become part of the general field pulsar population alongside the dispersed stellar members, making their natal association difficult to establish.  Secondly, observational challenges of detecting pulsars in extended tidal structures may have played a role. As our simulations demonstrate and the location of PSR~J1921$+$3745 exemplifies, NSs receiving low natal kicks can be retained within their host cluster for Gyrs and subsequently be stripped into tidal tails as the cluster evolves. If previous OC pulsar surveys focused primarily on the dense main bodies of the clusters, a substantial population of NSs residing in the dynamically evolving, extended tidal features could have been overlooked.  

Although our N-body simulations suggest that dozens of neutron stars could remain gravitationally bound to NGC 6791 or reside within its extended tidal tails, the probability of detecting radio emission from any individual neutron star after several Gyrs is still intrinsically low. Only $\sim$10\% of neutron stars are observable as radio pulsars due to beaming geometry constraints~\cite{Tauris1998, Kolonko2004}. Canonical pulsars remain radio-active for only a few million years. Accordingly, in the ATNF catalog$^{\ref{ATNF}}$ only $\sim$15\% of known pulsars have characteristic ages greater than 1 Gyr, and fewer than $\sim$5\% are older than the 8 Gyr age of NGC 6791. The rarity of radio detections is therefore consistent with expectations, and PSR J1921+3745 likely represents the detectable tip of a much larger, predominantly invisible neutron-star population associated with the cluster.

\subsection{ECSN NS retention with alternative kick dispersion and binary-orbit effects} 
\label{sec:ecs_ns_binary_retention}

In the simulation, the ECSNe kick velocities have a Maxwellian kick dispersion of  $\sigma$ = 3.0 pc$\,Myr^{-1}$(corresponding to a median velocity of $\simeq 4.2\,{\rm km\,s^{-1}}$). 
The low ECSN kicks follow the default low-kick ECSN prescription used in our \petar~ code and the Nbody6++GPU series of  $N$-body simulations\cite{Wang2020b, Kamlah2022}. 
This low-kick ECSN prescription is motivated by hydrodynamical simulations of ECSNe \cite{Gessner2018}. Based on 2D and 3D ECSN simulations, they found that the resulting NS kick velocities are only a few ${\rm km\,s^{-1}}$ at most, because of the weak ejecta asymmetry in ONeMg-core collapse.

We note that the ECSN kick distribution remains uncertain,  different studies adopt different ECSN kick prescriptions, including larger values such as $\sigma \sim 15$--$30\,{\rm km\,s^{-1}}$.
 We therefore perform a post-processing sensitivity test to quantify how the retention of ECSN-formed NSs in an NGC 6791-like cluster depends on the assumed kick dispersion and on possible binary-orbit effects.
 Here below we summarize the calculation methods, and list the NS retention rate in Table~\ref{tab:ecs_ns_final_retention}.

\subsubsection{Sampling of kick velocity}

Based on the main $N$-body simulation, we evaluate ECSN NS retention with different kick velocities using a Monte Carlo framework as described below.

Let $\{\mathbf{r}_{i,0},\mathbf{v}_{i,0}\}_{i=1}^{N_{\rm ECSN}}$ be the position and velocity of ECSN NSs at the first NS record used in this analysis.
Each NS has a record of the original SN-kick magnitude $v_{{\rm k},i}^{\rm old}$.
We remove the old kick amplitude along the current velocity direction:

\begin{equation}
\hat{\mathbf{v}}_{i,0}=\frac{\mathbf{v}_{i,0}}{\|\mathbf{v}_{i,0}\|},\qquad
\mathbf{v}_{i,{\rm base}}=\mathbf{v}_{i,0}-v_{{\rm k},i}^{\rm old}\,\hat{\mathbf{v}}_{i,0}.    
\end{equation}

Then, we draw a new 3D isotropic kick vector by sampling Cartesian components with 1D dispersion $\sigma$:
$v_{{\rm k},x},v_{{\rm k},y},v_{{\rm k},z}\sim\mathcal{N}(0,\sigma^2)$, where $\mathcal{N}(0,\sigma^2)$ is a normal distribution with mean $0$ and variance $\sigma^2$.
The kick-speed magnitude $v_{\rm k}=\sqrt{v_{{\rm k},x}^2+v_{{\rm k},y}^2+v_{{\rm k},z}^2}$ therefore follows a Maxwell distribution.
The post-replacement velocity is

\begin{equation}
\mathbf{v}_{i,{\rm new}}=\mathbf{v}_{i,{\rm base}}+\mathbf{v}_{{\rm k},i}.    
\end{equation}

For each object, the local escape speed is evaluated at $(t_{i,0},r_{i,0})$ by
\begin{equation}
 v_{{\rm esc},i}=\sqrt{\frac{4GM_{\rm enc}(r_{i,0},t_{i,0})}{r_{i,0}}},    
\end{equation}
where $M_{\rm enc}(r_{i,0},t_{i,0})$ is the enclosed mass within $r_{i,0}$ at time $t_{i,0}$.

To evaluate $M_{\rm enc}(r_{i,0},t_{i,0})$, we first use the nearest Lagrangian radii at $t_{i,0}$. The Lagrangian radii represent spherical radii enclosing a given mass fraction $f_{\rm m}$ of the cluster, measured from the density center. We calculate the radii with Lagrangian mass-fraction bins $f_{\rm m}=[0.1,\,0.3,\,0.5,\,0.7,\,0.9]$.
Then, for each NS, we linearly interpolate between the two nearest Lagrangian radii around $r_{i,0}$ at $t_{i,0}$ to estimate $M_{\rm enc}(r_{i,0},t_{i,0})$.
At the boundaries, interpolation uses constant extrapolation: below the innermost bin, $M_{\rm enc}=M_{\rm enc}(r_{\min})$; above the outermost bin, $M_{\rm enc}=M_{\rm enc}(r_{\max})$.

An ECSN NS is counted as retained if $\|\mathbf{v}_{i,{\rm final}}\|<v_{{\rm esc},i}$,
where $\mathbf{v}_{i,{\rm final}}$ is the final velocity after kick replacement (and binary treatment, if applicable).

We adopt $\sigma\in\{0,\,3,\,5,\,10,\,30\}\ {\rm pc\,Myr^{-1}}$, with $N_{\rm trial}=200$ realizations for each $(\sigma,\text{orbit scenario})$.
For any reported fraction $f$, we store trial values $\{f_t\}$ and report
\begin{equation}
\bar f=\frac{1}{N_{\rm trial}}\sum_{t=1}^{N_{\rm trial}} f_t,\qquad
s_f=\sqrt{\frac{1}{N_{\rm trial}}\sum_{t=1}^{N_{\rm trial}}(f_t-\bar f)^2}.    
\end{equation}
Error bars correspond to trial-to-trial scatter ($\pm s_f$), not the standard error of the mean.

\subsubsection{Sampling of binary orbits}

The simulation does not include primordial binaries. To investigate whether binary orbits can affect the escape rate of ECSN NSs, we evaluate both assumptions for each NS: binary and non-binary. 

Under the binary assumption, because the cluster is old, low-mass companions are expected to dominate surviving binaries. Therefore, we adopt a companion-mass range of $m_2\in[0.1,1.2]~M_\odot$.
ECSN-like events are expected to have relatively small ejecta masses compared with canonical CCSNe, so we adopt a moderate mass-loss range $\Delta M\in[0.1,0.8]~M_\odot$ as a phenomenological bracket.
Log-uniform sampling in $a$ is used as a scale-free prior over orders of magnitude in separation.
We separate $a$ into three sets (close/fiducial/wide) to evaluate the influence of pre-SN orbital hardness from compact to loose binaries.
For eccentricity, a low-$e$ prior is used for the close case to mimic partial tidal circularization, while the thermal prior $p(e)=2e$ is used for fiducial/wide cases as a standard baseline for dynamically mixed binaries.
Pre-SN binary parameters are sampled for every ECSN NS and every trial as:

\begin{enumerate}
\item Companion mass: $m_2\sim U(0.1,1.2)\,M_\odot$, where $U(a,b)$ is a uniform distribution on $[a,b]$.
\item SN mass loss: $\Delta M\sim U(0.1,0.8)\,M_\odot$, and the pre-SN primary mass is set to $m_{1,{\rm pre}}=m_{\rm NS}+\Delta M$.
\item Semi-major axis: log-uniform sampling, $\log_{10}a\sim U(\log_{10}a_{\min},\log_{10}a_{\max})$.
\item True anomaly: $f\sim U(0,2\pi)$.
\item Orbital orientation: isotropic random orientation in 3D.
\end{enumerate}

Orbit scenarios are defined by $(a_{\min},a_{\max})$ and eccentricity prior:

\begin{itemize}
\item close: $a\in[0.02,0.5]$ AU, $e\sim U(0,0.3)$;
\item fiducial: $a\in[0.05,5.0]$ AU, $e\sim p(e)=2e$ on $[0,1)$ (thermal);
\item wide: $a\in[0.5,50.0]$ AU, $e\sim p(e)=2e$ on $[0,1)$ (thermal).
\end{itemize}

Using the sampled $(a,e,f)$ and masses, we construct pre-SN relative motion, add the replacement kick, and compute the post-SN specific orbital energy:
\begin{equation}
  \epsilon_{\rm post}=\frac{1}{2}\|\mathbf{v}_{\rm rel,post}\|^2-\frac{G(m_{\rm NS}+m_2)}{r}.
\end{equation}
Binary survival is defined by $\epsilon_{\rm post}<0$.

For disrupted systems, the NS final velocity is taken as the pre-SN stellar velocity plus kick; for surviving systems, it is taken as the post-SN binary center-of-mass recoil velocity. Using this $\mathbf{v}_{i,{\rm final}}$, we evaluate cluster retention against the local escape speed.

\subsubsection{ECSN NS retention for different kick and binary assumptions}

For the $N_{\rm ECSN}=96$ we obtained from the baseline simulation, their instantaneous retention fractions and final survival rate at the cluster's age 8 Gyr,   with different kick velocities and binary assumption are listed in Table~\ref{tab:ecs_ns_final_retention}.

These post-processing tests are not full re-runs of the cluster simulation with primordial binaries; rather, they are designed to isolate how natal kicks and possible pre-SN binary orbital velocities affect the immediate retention probability of ECSN NSs.

The key trends are as follows:
\begin{itemize}
    \item  Kick-dispersion dependence. 
Cluster NS retention decreases strongly with higher kick velocity (increasing $\sigma$).

    \item Binary-orbit effects.
    At low $\sigma$ ($\lesssim 5\,{\rm pc\,Myr^{-1}}$), the binary assumption usually gives lower retention than the non-binary assumption in the close and fiducial cases, while the wide case can remain comparatively more permissive because many binaries are soft and their retained subset is dominated by systems that receive favorable recoil velocities. At high $\sigma$ ($\gtrsim 10\,{\rm pc\,Myr^{-1}}$), the difference between assumptions is small.
\end{itemize}

These tests demonstrate that the retention of ECSN NSs is primarily controlled by the natal-kick amplitude. For low kick dispersions of a few km s$^{-1}$, a substantial fraction of ECSN NSs can remain associated with the cluster even at 8 Gyr. However, when the dispersion is increased to $\sigma=30\,{\rm pc\,Myr^{-1}}(\simeq29\,{\rm km\,s^{-1}})$, the retained fraction falls to the percent level in both the non-binary and binary-orbit experiments. Therefore, the existence of PSR~J1921$+$3745 in the tidal field of NGC~6791  favors a low-kick formation channel.

\section{The radio properties of PSR~J1921+3745}
\begin{figure*}
    \centering
    \includegraphics[width=0.95\linewidth]{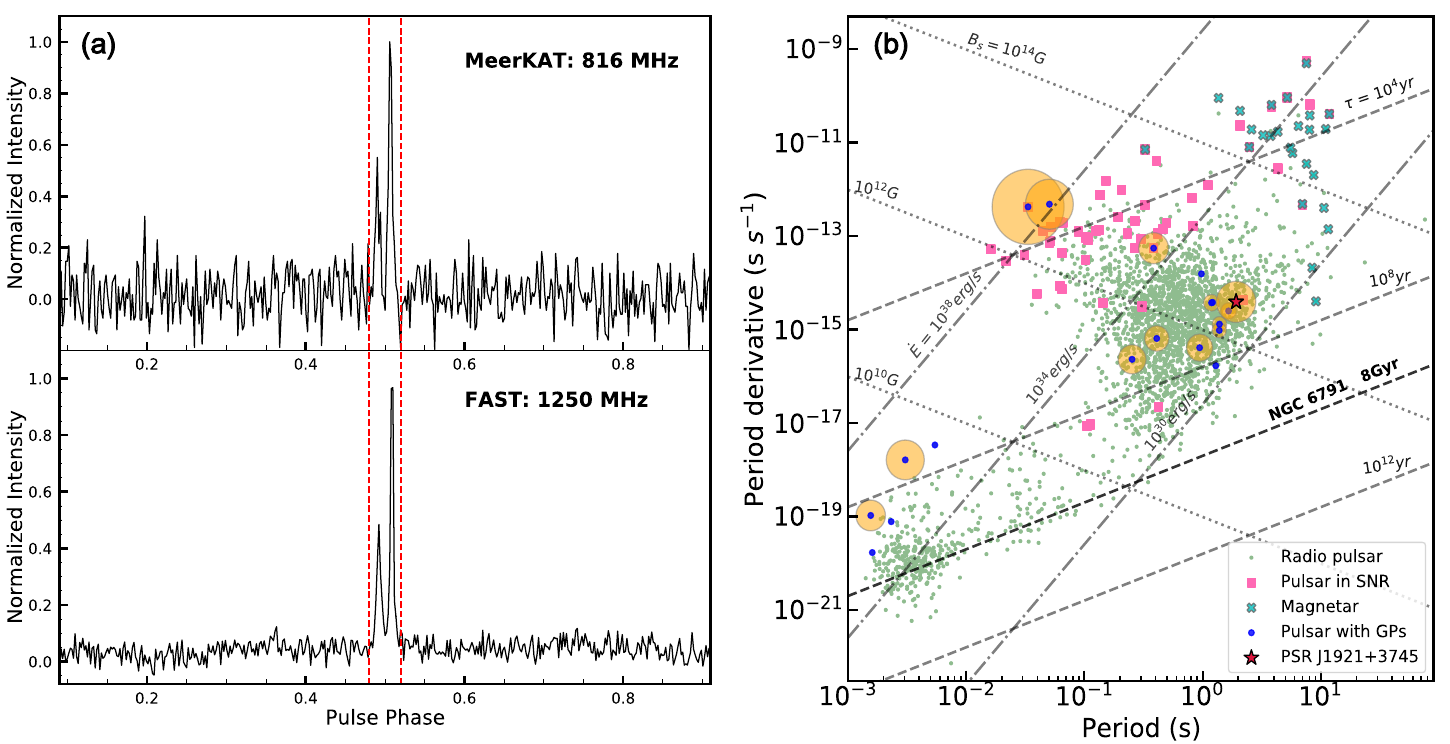}
    \caption{\small {\bf (a) The integrated pulse profile of PSR~J1921+3745.} Top panel: The result of the strongest detection with the MeerKAT UHF band, about 2.0\,hr integration time. Bottom panel: The result of the strongest detection with the FAST L-band, about 1.0\,hr integration time. The red vertical dashed line marks the width of the averaged double-peaked pulse profile. 
    {\bf (b) $P-\dot{P}$ diagram based on the Australia Telescope National Facility pulsar catalogue.} Dotted and dashed light grey lines represent constant magnetic field strengths (in Gauss) and characteristic ages (in years), respectively. Dot-dashed light grey lines correspond to constant rotational energy loss rates (in erg s$^{-1}$). The position of PSR~J1921$+$3745 is based on the recently reported timing solution, corresponding to a characteristic age of $\sim$7.8 Myr~\cite{LiuXJ2026}. The pulsar sub-classes are represented by the markers in the legend (SNR: supernova remnant; GPs: giant pulses). The size of each orange circle is proportional to the brightness factor of the giant pulses, ranging from 65 to 50,000 \cite{Malov22}. In comparison, this factor is $\sim$2000 for PSR~J1921$+$3745.
    }
    \label{fig_J1921}
\end{figure*}

PSR~J1921+3745 exhibited an estimated mean flux density of $S_{\text{FAST}} = 6.7 \pm 1.1$\,$\mu$Jy in the FAST L band and $S_{\text{MeerKAT}} = 8.0 \pm 1.4$\,$\mu$Jy in the MeerKAT UHF band. A summary of the measured parameters for PSR~J1921$+$3745 is provided in Table~\ref{tab:J1921_para}. The integrated pulse profiles at 816 MHz (MeerKAT) and 1250 MHz (FAST), shown in Fig.~\ref{fig_J1921}a, exhibit a narrow double-peaked morphology. The pulse width at 50\% of the peak intensity (W${50}$) is $\sim 11$\,ms, corresponding to a duty cycle of $\sim0.6\%$. Red dashed lines mark the pulse width at 10\% of the peak intensity (W${10}$).
Notably, no evidence of interstellar scattering was observed, even in the MeerKAT low-frequency band, with the scattering timescale constrained to $<0.2$\,ms. 

\begin{figure*}[]
    \centering
    \includegraphics[height=8.2cm,width=5.2cm]{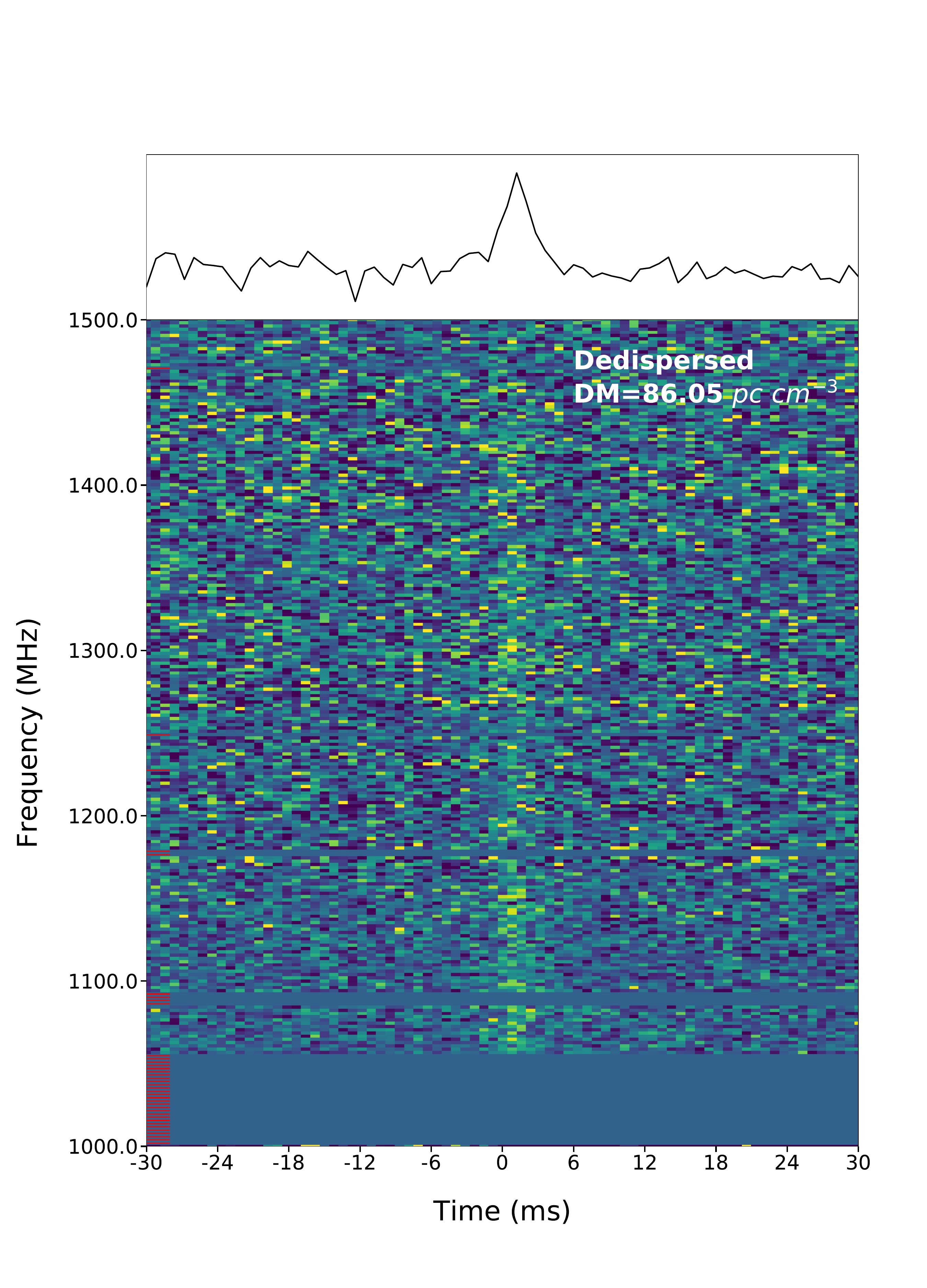}
    \includegraphics[height=8.2cm,width=5.2cm]{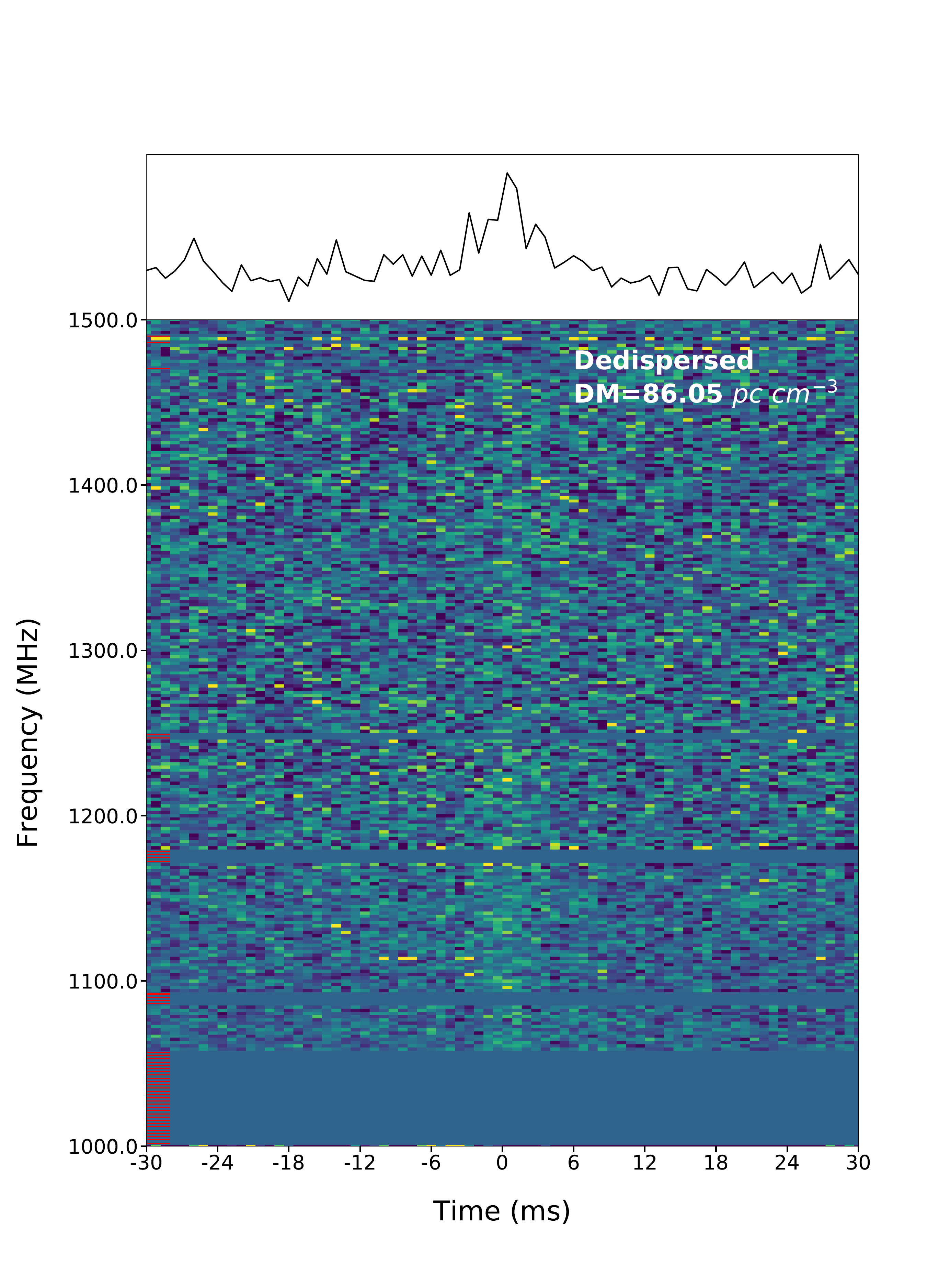}
    \includegraphics[height=8.2cm,width=5.2cm]{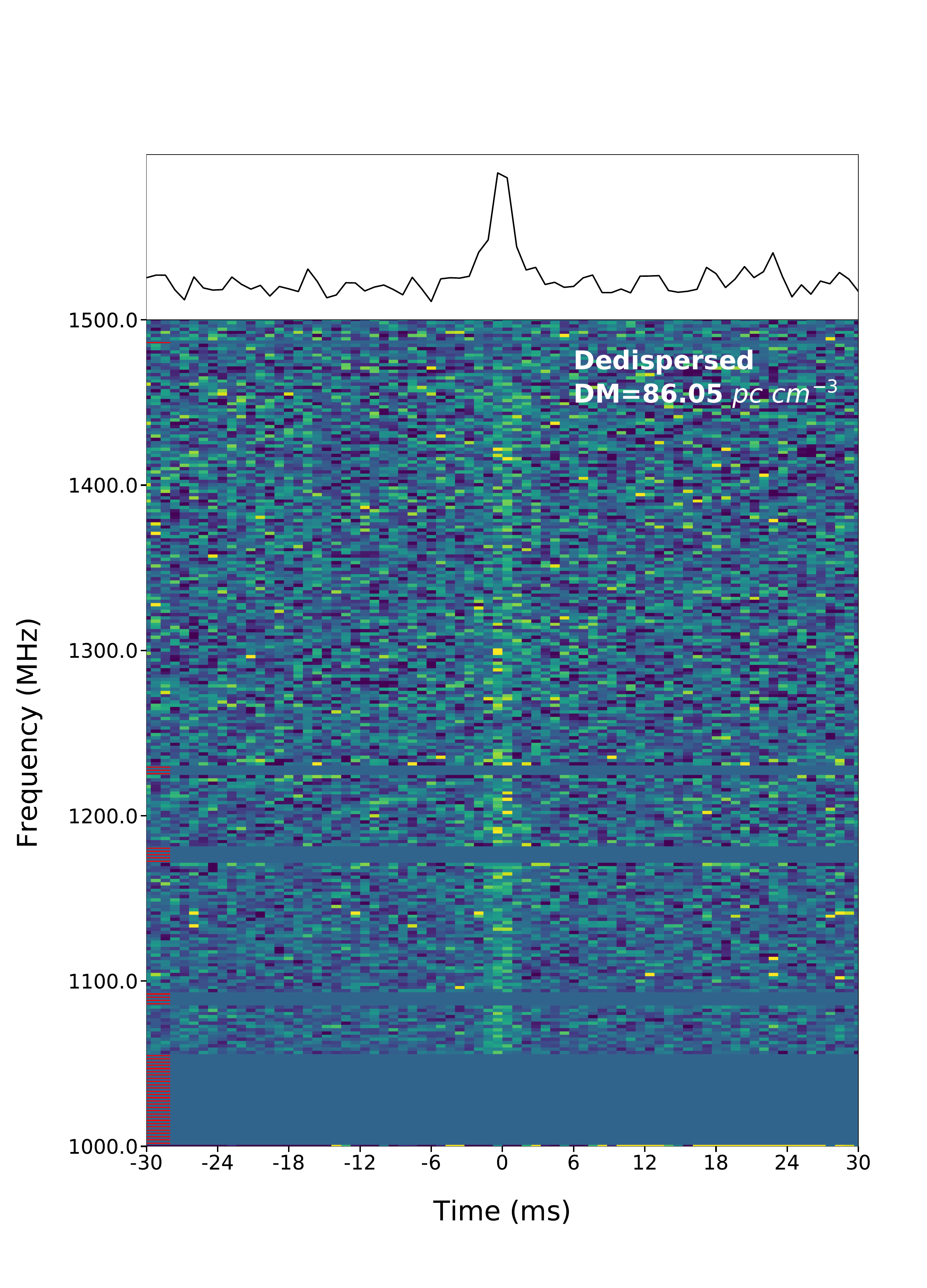}
    \caption{ {\bf Bursts from PSR~J1921+3745, shown as dynamic spectra and pulse profiles.} Top: frequency-integrated pulse normalized intensities that were detected in 1\,hour observation with FAST data done on 2022-01-22. Bottom: de-dispersed dynamic spectra of the pulses. The color map is linearly scaled and brighter patches represent higher intensities. The bad frequency channels are masked and labelled using red patches on the left. The time and frequency resolutions for the plots are downsampled to 0.786\,ms and 1.95\,MHz, respectively.}   
    \label{fig_SPs-FAST}
\end{figure*}

Additionally, a single-pulse search was performed across all beams during the same observing campaign. In FAST data collected on 22 January 2022, three bursts were detected (Fig.~\ref{fig_SPs-FAST}), with the strongest pulse exhibiting an S/N of $\sim 15$ and an estimated fluence of $\sim 80$\,mJy~ms. In MeerKAT data collected on 17 December 2024, 35 bursts were detected with a similar fluence and S/N of about 10. The properties of the MeerKAT bursts are presented in Table ~\ref{tab:MKT-SPs}. 

%From the spin period measurements obtained with FAST and MeerKAT (Table~\ref{tab:J1921_para}), we estimate an upper limit on the period derivative of PSR~J1921$+$3745, $\dot{P} < 1.015 \times 10^{-13} \text{s s}^{-1}$, which implies a characteristic age lower limit of $\tau_c = P/2\dot{P}  >$ 0.3\,Myr. The lack of nearby massive-star formation region excludes a young-magnetar origin, as discussed in the main text. An ``interrupted-recycling'' explanation for the long spin period is also unlikely: even globular clusters are not dense enough to halt mass transfer quickly to produce the $P$-$\dot{P}$ combination. The exact $\dot{P}$ value of PSR~J1921$+$3745 under its upper limit does not change the dynamical picture. 

Fig.~\ref{fig_J1921}b presents the $P-\dot{P}$ diagram for PSR~J1921+3745 alongside 18 other known pulsars exhibiting giant pulses~\cite{Malov22}. With a spin period of 1.9 seconds, PSR~J1921$+$3745 is the slowest-spinning pulsar yet observed to emit giant pulses. The giant pulses detected in MeerKAT's beam have a single pulse S/N of 10, while the total folded S/N of the observation, which contains 3736 pulses is about 20. This give a normal pulse S/N of about 0.005. Therefore, the giant pulses are 2000 times stronger than a normal pulse. A similar calculation for the FAST observation gives a factor of $\sim$1300. 

Giant pulses are sporadic flare-like events that were first discovered in the main pulse and inter-pulse of the Crab pulsar and are characterized by their relatively high intensities ($\lesssim10^{15}$\,erg s$^{-1}$), short durations (up to a few nanoseconds), and the very presence of circular polarization (of both signs) \cite{Hankins2007}. To date, giant pulses have been observed in 18 canonical and millisecond pulsars, and they show a power-law distribution, unlike the exponential or Gaussian distribution that normal pulses obey \cite{Malov2023}. Thoroughly investigating the origin of giant pulses is crucial for understanding the differences between pulsars that exhibit giant pulses and those that do not, and whether there is an age dependency. Although several models have been proposed in the literature to explain the giant pulses, the emission mechanisms have still not been fully elucidated. In the pulsar magnetosphere, at different distances from the surface, the proposed mechanisms include: (i) resonator wave guide in the vacuum gap \cite{Kontorovich2010}, (ii) electron turbulence \cite{Weatherall2001} or induced scattering at moderate distances \cite{Petrova2004}, and (iii) magnetic reconnection \cite{Istomin2004}, plasma instabilities \cite{Wang2019} or resonant interaction of plasma waves \cite{Lyutikov2007,Machabeli2019} near the light cylinder. In particular, the electric discharge at light cylinder due to the magnetic reconnection of the field lines connecting the magnetic poles held at opposite voltages was suggested as a plausible mechanism for explaining many different characteristics of giant pulses seen from the Crab pulsar and the millisecond pulsar PSR B1937+21 \cite{Istomin2004}. According to this model, the magnetic field at the light cylinder $B_{\rm LC}$ should be large enough. If we assume that this is the relevant mechanism for producing the giant pulses of PSR~J1921$+$3745 and adopt the lowest value of $B_{\rm LC}\gtrsim1$ G inferred for the sample in \cite{Malov22}, then we obtain the limit $\dot{P} > 7.51 \times 10^{-17} \text{s s}^{-1}$. This model is applicable when the pulsar has a large magnetic inclination angle $\alpha$ between its rotation and dipole moment axes. From the $W_{10}$ and $W_{50}$ values in Table~\ref{tab:J1921_para} we obtain the estimates for $\alpha$ as $48^\circ$ \cite{Kenko2023} and $66^\circ$ \cite{Rankin1990}, respectively. The gravitational potential field of the open cluster or  accretion/propeller interactions with cluster member stars, may have led to an increase in the magnetic inclination angle in such a potentially old pulsar by driving magnetic flux tubes radially out of its interior \cite{Chen1998}.

\section{Extended data tables and figures }
\clearpage
% === Final ECS NS retention table for model m100000_r4/with_tid ===
\begin{table*}[t]
\centering
\caption{Final ECSN NS retention at  8 Gyr under different kick-dispersion $\sigma$ and binary-orbit assumptions. $f_{\rm inst}$ is the instantaneous retention fractions just after the ECSN. $N_{\rm final}$ and $f_{\rm final}$ are the  retained NS number at 8 Gyr and the corresponding final retention fractions.
The unit pc Myr$^{-1}$ is nearly identical to km s$^{-1}$: $1\,{\rm pc\,Myr^{-1}}=0.978\,{\rm km\,s^{-1}}$.}
\label{tab:ecs_ns_final_retention}
\begin{tabular}{lcccc}
\hline
$\sigma$ [pc\,Myr$^{-1}$] & Scenario & $f_{\rm inst}$ & $f_{\rm final}$ & $N_{\rm final}$ \\
\hline
0.0 & non-binary & $1.000$ & $0.501$ & $48$ \\
0.0 & close & $0.560$ & $0.281$ & $27$ \\
0.0 & fiducial & $0.553$ & $0.277$ & $26$ \\
0.0 & wide & $0.752$ & $0.377$ & $36$ \\
\hline
\textbf{3.0} & \textbf{non-binary} & $\textbf{0.894}$ & $\textbf{0.448}$ & $\textbf{43}$ \\
3.0 & close & $0.513$ & $0.257$ & $24$ \\
3.0 & fiducial & $0.499$ & $0.250$ & $24$ \\
3.0 & wide & $0.659$ & $0.331$ & $31$ \\
\hline
5.0 & non-binary & $0.627$ & $0.314$ & $30$ \\
5.0 & close & $0.433$ & $0.217$ & $20$ \\
5.0 & fiducial & $0.402$ & $0.201$ & $19$ \\
5.0 & wide & $0.508$ & $0.255$ & $24$ \\
\hline
10.0 & non-binary & $0.183$ & $0.092$ & $8$ \\
10.0 & close & $0.223$ & $0.112$ & $10$ \\
10.0 & fiducial & $0.180$ & $0.090$ & $8$ \\
10.0 & wide & $0.199$ & $0.100$ & $9$ \\
\hline
30.0 & non-binary & $0.009$ & $0.005$ & $0$ \\
30.0 & close & $0.025$ & $0.013$ & $1$ \\
30.0 & fiducial & $0.013$ & $0.007$ & $0$ \\
30.0 & wide & $0.012$ & $0.006$ & $0$ \\
\hline
\end{tabular}
\par\vspace{4pt}\noindent{\footnotesize Notes: 
%$f_{\rm inst}$ = instantaneous post-SN survival fraction from Monte Carlo (200 trials). 
`non-binary' assumes all ECS NS are single; `close'/`fiducial'/`wide' assume all ECSN NS are in primordial binaries with the specified orbital priors. Bold row: baseline simulation ground truth ($\sigma=3$, non-binary).}
\end{table*}

\begin{table*} % Do NOT use \begin{table*}
\centering
\caption{Parameters for PSR~J1921+3746.}
\label{tab:J1921_para}
\renewcommand{\arraystretch}{1.5}
\setlength{\tabcolsep}{1mm}{
\begin{tabular}{lcc} \\\hline
Parameter                       & FAST                     &  MeerKAT\\
                                & 1000--1500\,MHz           &  544--1088\,MHz \\\hline
\multicolumn{3}{l}{\bf Measured parameters:}\\
Right ascension (J2000)$^{a}$   & \multicolumn{2}{c}{$19^{h}\,21^{m}\,59^{s}.70\pm0^{s}.084$} \\
Declination (J2000)$^{a}$       & \multicolumn{2}{c}{$+37^{\circ}\,45^{'}\,50^{''}.3\pm3^{''}.3$}\\
Period (ms)$^{b}$               & 1918.088$\pm$0.0032      &  1918.0926$\pm$0.0015\\
DM (pc cm$^{-3}$)$^{b}$         & 85.2$\pm$2.6             &  85.9$\pm$1.6\\
Epoch (MJD)                     & 59601.17260              &  60661.50429\\
$W_{\text{50}}$ (ms)$^{c}$        & 10.3$\pm$3.7            &  11.5$\pm$3.7\\
$W_{\text{10}}$ (ms)$^{c}$        & 55.6$\pm$3.7            &  55.6$\pm$3.7\\
Mean flux density ($\mu$Jy)     & 6.7$\pm$1.1              &  8.0$\pm$1.4\\\hline
\multicolumn{3}{l}{\bf Derived parameters:}\\
Galactic longitude, $l$ ($^{\circ}$)     & \multicolumn{2}{c}{70.05}\\
Galactic latitude,  $b$ ($^{\circ}$)     & \multicolumn{2}{c}{10.70} \\
Estimated distance, $d_{\text{NE2001}}$(kpc)$^{d}$ & $4.79_{-0.15}^{+0.16}$ & $4.84_{-0.10}^{+0.10}$\\
Estimated distance, $d_{\text{YMW16}}$ (kpc)$^{d}$ & $8.94_{-0.45}^{+0.47}$ & $9.07_{-0.28}^{+0.29}$\\
Estimated distance, $d_{\text{NE2025}}$ (kpc)$^{d}$ & $6.49_{-0.25}^{+0.26}$ & $6.56_{-0.15}^{+0.16}$\\
\hline
\multicolumn{3}{l}{{$^{a}$} Parameter obtained from MeerKAT data only.}\\
\multicolumn{3}{l}{{$^{b}$} Obtained with \textsc{psrchive}'s \texttt{pdmp} tool applied on high-S/N \textsc{dspsr} folded observation data and}\\
\multicolumn{3}{l}{~~~ transformed topocentric period to the Solar system barycentre.}\\
\multicolumn{3}{l}{{$^{c}$} The pulse widths at 50\% and 10\% of the peak intensity.}\\
\multicolumn{3}{l}{{$^{d}$} Derived from the NE2001~\cite{Cordes2002}, YMW16~\cite{Yao2017} and NE2025~\cite{Ocker2026} Galactic free electron density model.}\\
\end{tabular}}
\end{table*}

% NE2001_85.2+-2.6(pc): 4794.8256, 4955.5726, 4641.3550
% YMW16_85.2+-2.6(pc): 8944.1406, 9423.5508, 8485.6768
% NE2025_85.2+-2.6(pc): 6488.8517, 6747.3851, 6242.4733

% NE2001_85.9+-1.6(pc): 4837.3523, 4936.6250, 4740.9182
% YMW16_85.9+-1.6(pc): 9071.1455, 9367.1562, 8783.0723 
% NE2025_85.9+-1.6(pc): 6557.2756, 6716.6583, 6402.4538
\begin{table}
\centering
\caption{The properties of the pulses detected during the 2-hour MeerKAT observation on 2024-12-17. The Pulse Width is the width of the burst at 50\% of the peak intensity. The fluence values were calculated using the pulse S/Ns and widths reported by \textsc{transientx}.}
\label{tab:MKT-SPs}
\small
\renewcommand{\arraystretch}{1.7}
\setlength{\tabcolsep}{1.0mm}{
\begin{tabular}{cccccccc}\hline
No. & MJD  & Pulse Width & Fluence  & No. & MJD  & Pulse Width & Fluence \\
    &      & (ms)        & (mJy ms) &     &      & (ms)        & (mJy ms)\\\hline
1  & 60661.46549891769 & 3.86 & 63.30 & 19 & 60661.49389374610 & 7.71  & 47.33 \\
2  & 60661.46685316327 & 5.78 & 66.42 & 20 & 60661.49429337381 & 3.86  & 71.93 \\
3  & 60661.46714177956 & 3.86 & 92.07 & 21 & 60661.49482618846 & 3.86  & 82.72 \\
4  & 60661.46809641974 & 7.71 & 53.44 & 22 & 60661.49662445739 & 11.57 & 39.88 \\
5  & 60661.47471226287 & 3.86 & 74.09 & 23 & 60661.49671324867 & 3.86  & 69.05 \\
6  & 60661.47577789215 & 3.86 & 76.25 & 24 & 60661.49964374034 & 3.86  & 75.53 \\
7  & 60661.47651053181 & 3.86 & 64.02 & 25 & 60661.50614857207 & 5.78  & 61.13 \\
8  & 60661.48201632771 & 3.86 & 89.20 & 26 & 60661.51127698270 & 3.86  & 61.14 \\
9  & 60661.48279331839 & 9.64 & 45.97 & 27 & 60661.51660515142 & 3.86  & 88.48 \\
10 & 60661.48350373791 & 5.78 & 51.73 & 28 & 60661.52117850497 & 5.78  & 61.72 \\
11 & 60661.48561282096 & 5.78 & 67.60 & 29 & 60661.52222195874 & 5.78  & 58.19 \\
12 & 60661.48565721660 & 5.78 & 54.67 & 30 & 60661.52646230034 & 3.86  & 84.16 \\
13 & 60661.48738891438 & 3.86 & 64.02 & 31 & 60661.53174611804 & 5.78  & 61.13 \\
14 & 60661.48787733338 & 3.86 & 87.04 & 32 & 60661.53179051368 & 3.86  & 62.58 \\
15 & 60661.48883195124 & 5.78 & 57.02 & 33 & 60661.53421041085 & 3.86  & 83.44 \\
16 & 60661.48929815011 & 5.78 & 49.96 & 34 & 60661.53429922445 & 3.86  & 75.53 \\
17 & 60661.49311671081 & 5.78 & 64.07 & 35 & 60661.54342375603 & 5.78  & 68.19 \\
18 & 60661.49364954776 & 3.86 & 88.48 &    &                   &       &       \\\hline
\end{tabular}}
\end{table}
\begin{table}
    \centering
    %\caption{Present-day position and velocity of NGC~6791 center. $\alpha$, $\delta$ are Ra and Dec, $v_{\text{r}}$ is radial velocity, $\mu$ denotes proper motions, and $d$ is distance.}
    \caption{Present-day position and motion of the NGC 6791 cluster. Right ascension ($\alpha$), declination ($\delta$), radial velocity ($v_{\text{r}}$), proper motions ($\mu$), and distance ($d$) of the cluster center, used as initial conditions in our dynamical simulation.}
    \renewcommand{\arraystretch}{1.7}
    \begin{tabular}{cc}
        \hline
       $\alpha$  &  290.221 \\
       $\delta$  & 37.778 \\
       $v_{\text{r}}$ & $-$48.896 km/s \\
       $\mu_{\alpha\cos{\delta}}$ & $-$0.421 mas/yr \\
       $\mu_{\delta}$ & $-$2.269 mas/yr \\
       $d$ & 4231.0 pc\\
       \hline
    \end{tabular}
    \label{tab:init}
\end{table}

\clearpage

%\clearpage
%\bibliographystyle{model3-num-names}
%\bibliography{cas-refs}

\end{document}